\documentclass[a4paper,prl,superscriptaddress,aps,notitlepage,floatfix, twocolumn]{revtex4-2}
\usepackage[utf8]{inputenc}
\usepackage[T1]{fontenc}
\usepackage{amsmath, amssymb, array}
\usepackage[normalem]{ulem}
\usepackage[usenames]{color}
\usepackage{graphicx}
\usepackage{hhline,layout}
\usepackage{siunitx}
\usepackage{float}
\usepackage[pdftex,unicode,colorlinks,citecolor=blue,linkcolor=blue,
bookmarks=true,bookmarksopen=true,bookmarksopenlevel=3,bookmarksnumbered=true]{hyperref}

\definecolor{MKcolor}{rgb}{0.9, 0.15, 0}

\newcommand{\figref}[1]{Fig.~\ref{fig:#1}}

\begin{document}

\title{Mitigating quantum decoherence in force sensors by internal squeezing}
\author{M. Korobko}
\email{mkorobko@physnet.uni-hamburg.de}
\affiliation{Institut f\"ur Laserphysik und Zentrum f\"ur Optische Quantentechnologien der Universit\"at Hamburg,\\%
Luruper Chaussee 149, 22761 Hamburg, Germany}
\author{J. S\"udbeck}
\affiliation{Institut f\"ur Laserphysik und Zentrum f\"ur Optische Quantentechnologien der Universit\"at Hamburg,\\%
Luruper Chaussee 149, 22761 Hamburg, Germany}
\author{S. Steinlechner}
\affiliation{ Faculty of Science and Engineering, Maastricht University, Duboisdomein 30, 6229 GT Maastricht, The Netherlands}
\affiliation{ Nikhef, Science Park 105, 1098 XG Amsterdam, The Netherlands }
\author{R. Schnabel}
\affiliation{Institut f\"ur Laserphysik und Zentrum f\"ur Optische Quantentechnologien der Universit\"at Hamburg,\\%
Luruper Chaussee 149, 22761 Hamburg, Germany}
\date{\today}
\begin{abstract}
  The most efficient approach to laser interferometric force sensing to date uses monochromatic carrier light with its signal sideband spectrum in a squeezed vacuum state.
  Quantum decoherence, i.e. mixing with an
  ordinary vacuum state due to optical losses, is the main sensitivity limit.
  In this work, we present both theoretical and experimental evidence that quantum decoherence in high-precision laser interferometric force sensors enhanced with optical cavities and squeezed light injection can be mitigated by a quantum squeeze operation inside the sensor's cavity.
  Our experiment shows an enhanced measurement sensitivity that is independent of the optical readout loss in a wide range. 
  Our results pave the way for quantum improvements in scenarios where high decoherence previously precluded the use of squeezed light. 
  Our results hold significant potential for advancing the field of quantum sensors and enabling new experimental approaches in high-precision measurement technology.
\end{abstract}

\maketitle
\textit{Introduction}. Quantum-correlated light, such as squeezed light, has been successfully used to boost the sensitivity of a broad range of force sensors: from gravitational-wave detectors~\cite{LSC2011, Acernese2019, Tse2019, Yu2020, Lough2021}, optomechanical devices~\cite{Yap2020} and dark matter detectors~\cite{backesQuantumEnhancedSearch2021,Carney2021}, biological sensors~\cite{Taylor2013, Taylor2016} and magnetometers~\cite{Li2018}.
Reaching high sensitivity in these quantum-noise limited devices requires maximizing the carrier light's power to a level where the device remains operational.
A conventional method for accomplishing this is by employing optical cavities, which resonantly enhance both the power and signal.
The combination of cavity enhancement with quantum squeezed light allows achieving even higher sensitivity.
Yet, further improvement is hindered by quantum decoherence, caused by the loss of purity in squeezed state due to quantum vacuum being mixed in at lossy optical components.

The optimal design of quantum sensors relies on the ability to extract information out of the system.
This ability is quantified by dedicated metrics, such as quantum Cramer-Rao bound (QCRB), discovering new sensing approaches\,\cite{tsangQuantumTheorySuperresolution2016, bentleyDesigningOptimalLinear2023}. 
In an ideal lossless system, QCRB defines the best possible sensitivity of a device at every angular signal frequency $\Omega$ for a given optical power\,\cite{Braginsky2000, Tsang2011, Miao2017}. 
In reality, decoherence always prevents the (lossless) QCRB from being reached~\cite{Dorner2009,Demkowicz-Dobrzanski2013, Demkowicz-Dobrzanski2015, Miao2019}.
For example, the best phase sensitivity for a lossless Michelson interferometer for a given energy per measurement time, is achieved with \textit{N00N states}, where $N$ stands for the total number of photons\,\cite{Dowling2008}. 
However, it is known that quantum decoherence renders  \textit{N00N} states ineffective. 
Instead, the preferred method for achieving high sensitivity involves using monochromatic light combined with a vacuum state, squeezed in a relevant signal spectral band.\,\cite{Demkowicz-Dobrzanski2013}.
In this example, the use of a decoherence-induced quantum bound instead of a QCRB allows to find an efficient experimental design.
Such bounds are especially relevant when the loss in a sensor is high. 
The major source of decoherence in all cavity-enhanced sensors is the readout loss that occurs on the light after outcoupling from the cavity: e.g. due to propagation losses inside the waveguides of chip-based sensors or mode mismatch in the mode-filtered readout of a gravitational wave detector.
So far, readout loss has limited quantum enhancement from injected squeezing in force sensors.

Here, we show that quantum squeezing of the light inside the cavity-enhanced force sensor allows to mitigate a substantial portion of decoherence due to loss of optical quanta.
We experimentally mitigate up to 20\% of readout loss and show 4\,dB of sensitivity enhancement independent of the readout loss across a wide range.
Further, we prove theoretically, that our internal squeezing approach is the most effective strategy for boosting the sensitivity in the presence of loss.
We generalize previous studies of internal squeezing\,\cite{Rehbein2005, V.Peano2015, korobko2017beating, korobko2018engineering, korobko2019quantum,Miao2019,adyaQuantumEnhancedKHz2020} and establish the decoherence-induced limit for practical lossy cavity-enhanced sensors, demonstrating that this limit is ultimately dictated exclusively by the losses inside the detector cavity.
\begin{figure}
  \includegraphics*[width=0.9\columnwidth]{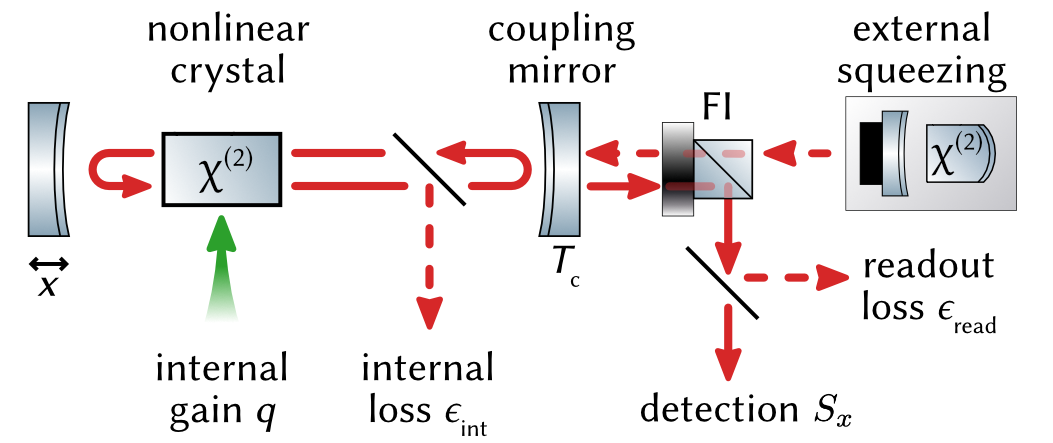}
  \caption{
  Conceptual representation of a cavity-enhanced sensor with internal and external squeezing.
  The detector cavity is used to measure the displacement $x$ of the movable end mirror. 
  This displacement imprints a phase modulation on the light field, which is amplified by the cavity.
  Internal squeezing is generated by a nonlinear crystal inside the cavity by pumping it with second harmonic pump. 
  Depending on the phase of the pump, the signal is either amplified or deamplified. 
  External squeezing is generated independently and is injected into the detector cavity through a Faraday isolator (FI).
  The output signal with external and internal squeezing $S_x$ is detected on a phase-sensitive detector.
  The impact of the readout loss $\epsilon_{\rm read}$ can be reduced by optimal choice of the internal gain $q$, defined by the pump strength.
  Optimal internal squeeze operation allows to reach the fundamental limit, defined by internal loss $\epsilon_{\rm int}$.
  }\label{fig:1}
\end{figure}

\textit{Internal squeezing for quantum sensors}.
In internal squeezing, a second-order nonlinear crystal is placed inside the cavity and pumped with second harmonic light. The pump phase determines which quadrature of the optical field is parametrically attenuated (squeezed), while the orthogonal quadrature is then amplified.
If the signal corresponds to a displacement of the quadrature being squeezed, it is accordingly deamplified.
Despite this, the sensitivity, defined by the signal-to-noise ratio (SNR) is still improved\,\cite{Rehbein2005, korobko2017beating}. 

A simple example for a force sensor is an optical cavity with a movable end mirror.
The external oscillating force displaces the movable mirror, creating a phase modulation signal on the reflected light field, which is detected with a balanced homodyne detector.
Here, we consider such a sensor employing quasi-monochromatic light with an externally injected squeezed sideband spectrum as well as a squeeze operation inside the sensor cavity (Fig.\,\ref{fig:1}). 
Optical loss in the system influences the sensitivity in different ways, depending on whether the loss occurs (i) {\it before} the measurement interaction, i.e.~on the meter alone, (ii) {\it after} the measurement, i.e.~on the meter that carries the full signal information, or (iii) {\it during} the time when the meter is accumulating the signal.
In the first case, the loss sets an upper bound on the external squeeze factor.
In the second case, the loss occurs after the information has been imprinted on the meter.
The third case is central to this work and more complex.
The loss occurs during the measurement process itself, thus changing the SNR on the out-coupled light.

We use the input-output formalism\,\cite{Caves1985a, Danilishin2012, Schumaker1985a} to derive the noise spectrum $S_{sn}(\Omega)$ and the power of signal transfer function $\mathcal{T}_x(\Omega)$, as well as the noise-to-signal ratio (sensitivity) $S_x(\Omega)$.
They take the form:
\begin{align}
& S_{\rm sn}(\Omega) = 1 - \frac{1-\epsilon_{\rm read}}{(T_c + \epsilon_{\rm int} + q)^2 + \Omega^2} \nonumber\\ &\times \left[4 T_c q + \left(1-\beta^{-1}\right) \left( (T_c -\epsilon_{\rm int} - q)^2 + \Omega^2\right)\right],\\
&|\mathcal{T}_x(\Omega)|^2 = \frac{8 \pi P_c}{\hbar \lambda c}\times\frac{T_c (1-\epsilon_{\rm read})}{(T_c + \epsilon_{\rm int} + q)^2 + \Omega^2},\\
&S_x(\Omega) = \frac{S_{\rm sn}(\Omega)}{|\mathcal{T}_x^{\rm}(\Omega)|^2}\label{eq:snr},
\end{align}
where $T_c$ is the coupling mirror power transmission, $\epsilon_{\rm int}$ is the internal power loss, and $q$ is the roundtrip parametric power gain, $\epsilon_{\rm read}$ is the readout power loss, see Fig.\,\ref{fig:1}, $\beta^{-1} = e^{-2r_{\rm ext}}$ is the external squeezing, with $r_{\rm ext}$ the corresponding squeeze parameter\,\cite{Stoler1970},  $c$ is the speed of light, $\lambda$ is the central wavelength and $P_c$ is intra-cavity light power. 
In deriving these equations we used the single-mode approximation, where only one optical mode acquires the signal, and $\{T_c, \epsilon_{\rm int}, q\}\ll1$\,\cite{korobko2023long}.
In this model we neglected quantum back-action effects such as photon radiation pressure, since they can be circumvented by quantum back-action evasion techniques \cite{Kimble2001}. 
The sensitivity of our sensor is limited only by quantum shot noise.
We fixed the average intra-cavity power $P_c$ and assumed no loss or depletion on the second harmonic pump power.

In order to simplify the conceptual explanation, we focus on the case of the peak sensitivity, which occurs at $\Omega=0$, and leave out the frequency dependence due to the cavity linewidth.
From the input-output relation in the lossless case we obtain the QCRB:
\begin{equation}
  S_{\rm QCRB}(0) = \frac{\hbar \lambda c}{8 \pi P_c} \times\frac{(T_c-q)^2}{\beta T_c},
\end{equation}
which turns to zero in the limit of infinite input squeezing $\beta\rightarrow\infty$ or at the (lossless) parametric threshold for internal squeezing $q = T_c$, resulting in the well-known theoretical limit of infinite SNR in lossless systems.
In \cite{Miao2019}, it was proposed that the loss-induced sensitivity limit is defined by the noise added to QCRB due to losses, when internal squeezing operates at threshold $q = q^{\rm th} := T_c + \epsilon_{\rm int}$.
However, it is possible to optimize internal squeezing to evade part of decoherence in the system and achieve even higher sensitivity.

Optimized internal squeezing enhances sensitivity in one of two ways:\\ (i) With low readout loss or external squeezing, it attenuates the signal quadrature.
The signal is maximally deamplified by 6\,dB, while quantum noise squeezing can be arbitrary high in the absence of loss\,\cite{Milburn1981,Collett1984,korobko2017beating}, increasing the overall SNR.
This increase is possible due to a different path for the signal and the noise in the sensor.
Maximal signal deamplification occurs at the pump power that corresponds to the optical oscillation threshold of the $\chi^{(2)}$ process.
In practice, there is an optimal parametric pump power value below threshold, which depends on the nonzero optical loss value.\\
(ii) With high readout loss or external squeezing, the nonlinear process amplifies both signal and quantum noise in signal quadrature.
In this case, the impact of the readout loss is mitigated, similarly to the output amplification discussed earlier in this section.
An important difference is that here amplification is realised inside the cavity, i.e.~already during the time when the signal is accumulated, which provides direct access to tuning the SNR of the out-coupled light.

The optimal internal cavity roundtrip gain $q$ depends on the quantities $\epsilon_{\rm int}$, $T_c$, $\epsilon_{\rm read}$, and $\beta$. 
It is computed by minimizing the value of the sensitivity $S_x(0)$ in Eq.\,\eqref{eq:snr}, which results in the optimal sensitivity:
\begin{align}
  S^{\rm opt}_{1/\beta}(0) &=\frac{\hbar \lambda c}{8 \pi P_c} \times 4 \left(\epsilon_{\rm int} +  \frac{T_c \epsilon_{\rm read}}{\epsilon_{\rm read}\beta + (1-\epsilon_{\rm read})}\right),\label{eq:optimal}\\
  q^{\rm opt} &= T_c \left(1 - \frac{2 \epsilon_{\rm read}}{\beta (1-\epsilon_{\rm read}) - \epsilon_{\rm read}}\right) - \epsilon_{\rm int}.\label{eq:optimal_gain}
\end{align}
The expression for $S^{\rm opt}_{1/\beta}$ is strictly lower than the lowest noise limit proposed in Ref.\,\cite{Miao2019} for non-zero readout loss.
A highly squeezed beam possesses substantial photon power. However, by assuming a suitably small measurement bandwidth, we can ensure that the power within this specific bandwidth remains small.
The fundamental sensitivity limit of our work is achieved for infinite external squeezing, $\beta\rightarrow\infty$.
In this case the optimal internal gain maximally amplifies the signal quadrature, $q = -q_{\rm th}$, and the limit becomes defined solely by the internal loss:
\begin{equation}\label{eq:limit}
  S^{\rm lim}_0(0) = 4\epsilon_{\rm int} \times \frac{\hbar \lambda c}{8 \pi P_c}.
\end{equation}
The equation formalizes the main theoretical result our paper: for a fixed power in the carrier field, if the external squeeze factor approaches infinity, the noise-to-signal ratio becomes \textit{independent on the readout loss} and approaches zero when the cavity internal loss approaches zero.
Injection loss limits the achievable level of external squeezing, thus the optimized sensitivity given by Eq.\eqref{eq:optimal_gain} represents the practical decoherence-induced limit.

\begin{figure}[t]
  \includegraphics[width=\linewidth]{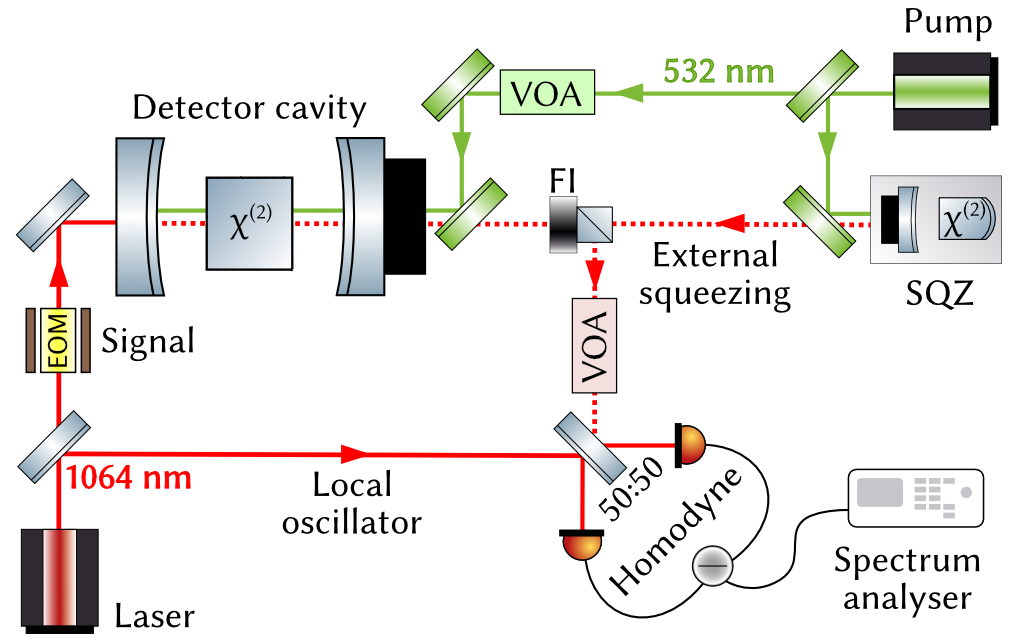}
  \caption{Schematic representation of the experimental setup.
  Phase modulation signal from the electro-optic modulator (EOM) was injected from the back of the detector cavity and was observed on the balanced homodyne detector by overlapping with local oscillator (LO). 
  The phase of the LO enable the selection of the quadrature of the light to observe.
  Pump light was used to create internal amplification in the detector cavity.
  The internal gain was adjusted by tuning the pump power with a variable optical attenuator (VOA).
  External squeezing was injected into the ISC through a Faraday Isolator (FI).
  The readout loss was adjusted with a VOA.}
  \label{fig:setup_exp}
\end{figure}

\textit{Experimental decoherence mitigation}. We demonstrated the mitigation of the readout loss and the existence of optimal internal squeezing in a table-top experimental setup, see \figref{setup_exp}.
Our internal squeezing cavity (ISC) was a Fabry-Perot cavity with a nonlinear periodically poled KTP crystal inside acting as an optical-parametric amplifier.
Parametric gain of $q$ was set by the power and the phase of the second harmonic pump.
The phase of the pump was actively controlled to keep the amplification phase stable.
\begin{figure}[ht!]
  \includegraphics[width=\linewidth]{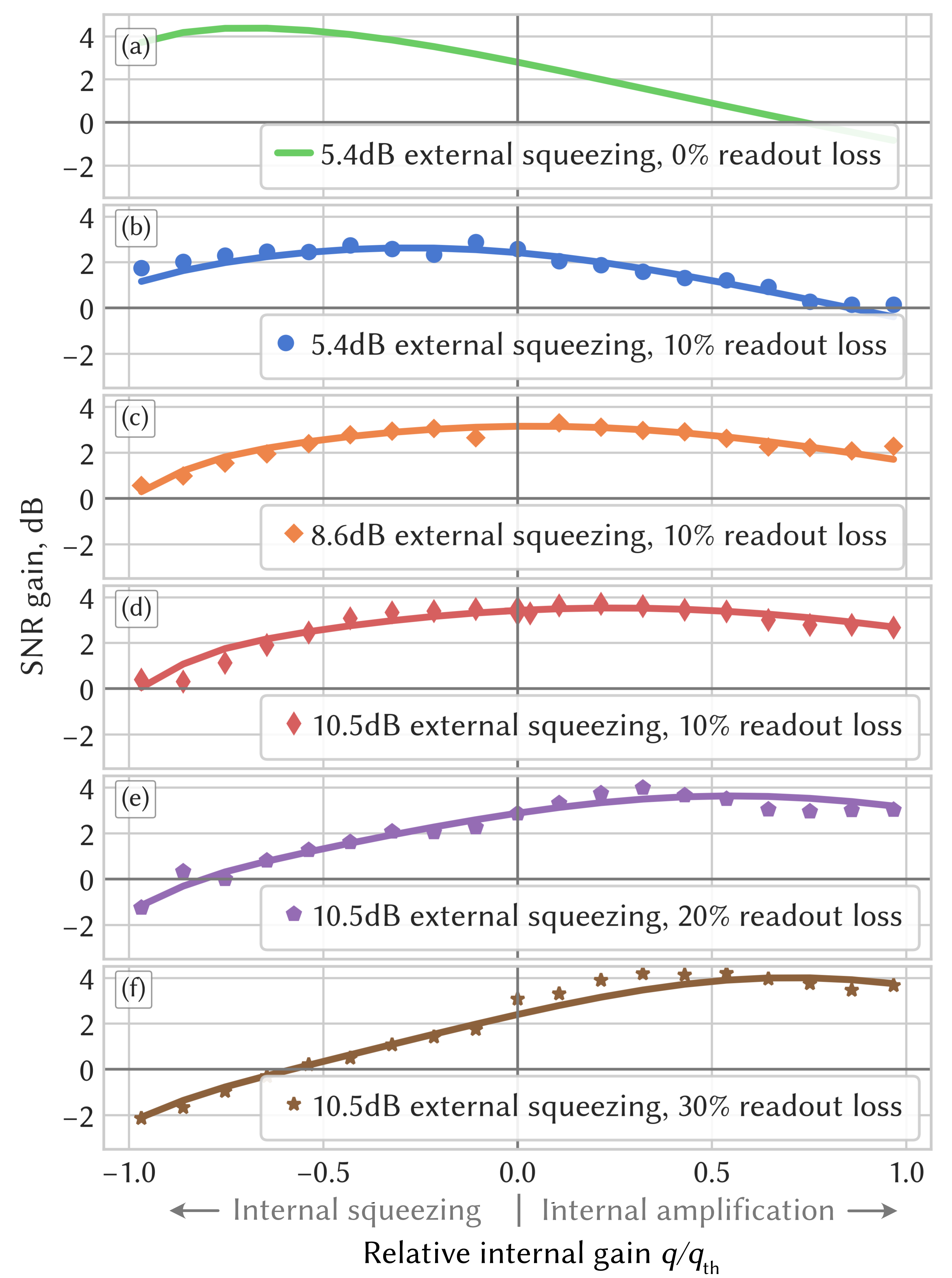}
  \caption{Demonstration of optimal sensitivity with internal and external squeezing approaching the optimal limit in Eq.\,\eqref{eq:optimal}.
  Each plot shows the improvement in the SNR with respect to the case without internal squeezing (with external squeezing).
  This improvement was observed as a function of internal gain in the detector cavity: negative gain means deamplification (squeezing), positive -- amplification.
  Solid curves are theoretical predictions based on the independently measured parameters.
Plot (a) shows a theoretical prediction for optimal operation in the absence of readout loss. Plots(b-d) demonstrate the different regimes of internal squeezing for different level of external squeezing: 5.4\,dB, 8.6\,dB and 10.5\,dB.
  For low values of external squeezing, peak SNR is achieved when squeezing is generated also internally.
  For high values of external squeezing it becomes optimal to amplify the signal quadrature.
  Plots (d-f) demonstrate the independence of the peak SNR enhancement on the readout loss.
  The enhancement of $\sim$4\,dB is achieved for 10\%, 20\% and 30\% of readout loss, for different levels of internal gain. 
  In all data sets, for the case of high squeezing (close to gain equal to $-1$)\,\cite{Schnabel2017}, the effect of phase noise played a significant role: due to the jitter in the phase of injected external squeezing, some part of anti-squeezed noise coupled into the readout quadrature, which further degraded the sensitivity (seen as strong degradation of the SNR close to gain equal to $-1$).
  The error bars on the experimental data are not shown, see the discussion in the main text.
  }
  \label{fig:full_exp}
\end{figure}
The cavity output field was analysed with a balanced-homodyne detector (BHD).
The phase of the BHD's local oscillator was actively controlled to keep the readout quadrature stable.
We injected a weak field carrying a 5 MHz phase modulation signal from the back of the ISC, which emulated the measurement signal.
Depending on the phase of the pump, we could observe amplification or deamplification of the signal, as well as anti-squeezing or squeezing of the noise.
By taking the spectrum of the signal and the noise, we could compute the change in the SNR compared to the case without the parametric process.
Further, we injected external squeezing from a squeeze laser\,\cite{Schnabel2017}.
We kept the external squeeze field without any bright carrier field and periodically varied its phase.
On the output of the homodyne detector, we recorded the spectrogram (spectrum as a function of time) of the observed signal.
As a result, we consecutively measured two orthogonal quadratures of the injected squeeze field, and used it to extract optical parameters of the setup:
the transmission of the incoupling mirror $T_c = 11$\%, the internal loss $\epsilon_{\rm int} = 1.2$\%, the injection loss of 8\% and the readout loss $\epsilon_{\rm read} = 10$\%\,\cite{Vahlbruch2016}.
Initial injected external squeezing was inferred to be 10.5\,dB. 
Phase noise of $\approx50$ mrad was inferred from the measurements with different pump strength\,\cite{Mehmet2011}.
We observed less phase noise for smaller values of external squeezing: 40 mrad for 8.6\,dB and 15 mrad for 5.4\,dB. 
We conclude that the main contribution came from the phase noise of the external squeezing interacting with the internal squeezing process.
We used these independently measured parameters to compute expected SNR gain based on our theoretical model.

We changed internal squeezing gain and recorded squeezing spectra together with amplified or deamplified signals.
In the first stage of the experiment, we gradually increased the injected external squeezing value from 5.4\,dB to 10.5\,dB.
When both the readout loss and the external squeezing were small, the optimal internal gain corresponded to an actual squeezing process, see Fig.\,\ref{fig:full_exp}(a-b).
As we increased external squeezing, the optimal internal gain approached zero, see Fig.\,\ref{fig:full_exp}(c), and then it became optimal to amplify inside the cavity, Fig.\,\ref{fig:full_exp}(d).

We further artificially increased the readout loss from 10\% to 30\%, by dumping part of the light on a polarizing beam splitter.
By taking the full range from maximal deamplification to maximal amplification we could find the optimal point where the SNR was the highest.
In the case of 10\% of readout loss, the optimal internal gain was close to zero, see Fig.\,\ref{fig:full_exp}(d).
For higher loss, as expected from our theory in Eq.\,\eqref{eq:optimal_gain}, it became optimal to \textit{amplify} the signal inside the cavity, see Fig.\,\ref{fig:full_exp}(e-f).
We observed the maximal sensitivity improvement of $\sim4$\,dB, independent of the readout loss.
This way, we demonstrated for the first time the mitigation of quantum decoherence.
Our results show good match with theory and demonstrate the significant enhancement over the case without decoherence mitigation (i.e.~with maximal internal squeezing), which was considered in\,\cite{Miao2019}.

We could not deduce meaningful error bars for Fig.~\ref{fig:full_exp}.
Most of the source data were averaged directly on the spectrum analyser, which did not allow for the extraction of variances.
However, since all data points are statistically independent, the good match between the data and the theoretically predicted SNR gain allows to be confident in the significance of our observed results in Fig.~\ref{fig:full_exp}.

\textit{Discussion and outlook}.
We have demonstrated for the first time how readout loss can be optimally mitigated in quantum-enhanced force sensors.
We used an internal squeezing approach that complemented injected external squeezing, reaching the newly established fundamental limit.
The ultimate value of this limit is determined exclusively by internal loss.
This loss is fundamentally different from the readout loss, since it occurs during the measurement process itself, thus changing the SNR on the out-coupled light. 
Any consequent quantum operation on this light would not be able to improve the SNR, therefore internal loss sets the fundamental limit.

Our result can be placed in more general context of studying the impact of state impurity on the QCRB\,\cite{tsang2013quantum,alipour2014quantum,fiderer2019maximal}, which has not been done for cavity-enhanced sensors.
While we do not derive the QCRB for our setup from the first principles, our argument follows the same spirit as Refs.\,\cite{tsang2013quantum,alipour2014quantum,fiderer2019maximal}.
There are two general conditions for achieving the QCRB: (i) the detector should be in a pure state, and (ii) the back-action of the meter should be evaded\,\cite{Tsang2011,Miao2017}.
The state's impurity not only prevents achieving an arbitrary low uncertainty in the signal quadrature, but also obstructs efficient back-action evasion, which relies on quantum correlations between the two quadratures\,\cite{Danilishin2012}. 
As a result, the primary constraint on attainable sensitivity is determined by the state's purity during the interaction between the meter and the object.
This is directly manifested in our fundamental limit in Eq.\,\eqref{eq:limit}, via its proportionality to cavity internal loss.

We also note that the output amplification proposed in Ref.\,\cite{Caves1981} also evades a part of the readout loss: the loss upon photodetection.
Both signal and noise are amplified, and thus the detrimental effect of vacuum mixed into the quantum state can become negligible.
This approach was recently explored in other contexts\,\cite{colomboTimereversalbasedQuantumMetrology2022, manceauDetectionLossTolerant2017, ouEnhancementPhasemeasurementSensitivity2012, knyazevOvercomingInefficientDetection2019,frascellaOvercomingDetectionLoss2021}.
In the output amplification approach, the signal is amplified \textit{after} it has exited the detector cavity.
Output amplification could only approach the fundamental limit if the inefficiency of the photodetection was the only readout loss.
In reality, the efficiency of the photodiodes can reach very high values\,\cite{Vahlbruch2016} and the main readout loss occurs at the outcoupling from the sensor.
This is the case, for example, with imperfect transverse mode shapes such as those found in gravitational-wave detectors and usually in chip-based optical waveguide sensors.
In the companion paper\,\cite{korobko2023long} we provide more details on the comparison between the two approaches.

Our results are readily applicable to force sensors that are limited by quantum shot noise, and whose principle schemes can be mapped to a single cavity.
This approach is especially promising for the cavities that naturally have nonlinear materials in them, such as on-chip devices\,\cite{Ramelow2019, Strekalov2016}, or whispering-gallery-mode sensors\,\cite{93a1BrGoIl, 94a1GoIl, Foreman2015, Schliesser2010}.
For these devices squeezing injection might be challenging, and the readout is often subject to significant losses.
There, internal squeezing can become a useful tool for mitigating these losses and achieving further quantum improvement to the sensitivity. 
Even in the systems with several cavities internal squeezing allows for the enhancement of the sensitivity, by quantum-expanding the linewidth\,\cite{korobko2019quantum,adyaQuantumEnhancedKHz2020} or creating PT-symmetric configurations\,\cite{dmitrievEnhancingSensitivityInterferometers2022, gardner2022nondegenerate}.
Our approach paves the way for implementing quantum enhancement in lossy experiments in which it was previously impractical.

The new fundamental sensitivity limit contributes to the detailed understanding of quantum limits on realistic sensors, and serves as a valuable guide for designing such experiments.
The use of optimal internal squeezing will enable the use of squeezed light in quantum technology beyond laboratory environments, opening new paths for technological progress.

\textit{Acknowledgements} This work was supported by the Deutsche Forschungsgemeinschaft (DFG) under Germany's Excellence Strategy EXC 2121 ``Quantum Universe''-390833306.

\bibliography{bibliographyCompensating}

\begin{thebibliography}{55}%
\makeatletter
\providecommand \@ifxundefined [1]{%
 \@ifx{#1\undefined}
}%
\providecommand \@ifnum [1]{%
 \ifnum #1\expandafter \@firstoftwo
 \else \expandafter \@secondoftwo
 \fi
}%
\providecommand \@ifx [1]{%
 \ifx #1\expandafter \@firstoftwo
 \else \expandafter \@secondoftwo
 \fi
}%
\providecommand \natexlab [1]{#1}%
\providecommand \enquote  [1]{``#1''}%
\providecommand \bibnamefont  [1]{#1}%
\providecommand \bibfnamefont [1]{#1}%
\providecommand \citenamefont [1]{#1}%
\providecommand \href@noop [0]{\@secondoftwo}%
\providecommand \href [0]{\begingroup \@sanitize@url \@href}%
\providecommand \@href[1]{\@@startlink{#1}\@@href}%
\providecommand \@@href[1]{\endgroup#1\@@endlink}%
\providecommand \@sanitize@url [0]{\catcode `\\12\catcode `\$12\catcode
  `\&12\catcode `\#12\catcode `\^12\catcode `\_12\catcode `\%12\relax}%
\providecommand \@@startlink[1]{}%
\providecommand \@@endlink[0]{}%
\providecommand \url  [0]{\begingroup\@sanitize@url \@url }%
\providecommand \@url [1]{\endgroup\@href {#1}{\urlprefix }}%
\providecommand \urlprefix  [0]{URL }%
\providecommand \Eprint [0]{\href }%
\providecommand \doibase [0]{https://doi.org/}%
\providecommand \selectlanguage [0]{\@gobble}%
\providecommand \bibinfo  [0]{\@secondoftwo}%
\providecommand \bibfield  [0]{\@secondoftwo}%
\providecommand \translation [1]{[#1]}%
\providecommand \BibitemOpen [0]{}%
\providecommand \bibitemStop [0]{}%
\providecommand \bibitemNoStop [0]{.\EOS\space}%
\providecommand \EOS [0]{\spacefactor3000\relax}%
\providecommand \BibitemShut  [1]{\csname bibitem#1\endcsname}%
\let\auto@bib@innerbib\@empty
\bibitem [{\citenamefont {{Abadie \emph{et al.}}}(2011)}]{LSC2011}%
  \BibitemOpen
  \bibfield  {author} {\bibinfo {author} {\bibfnamefont {J.}~\bibnamefont
  {{Abadie \emph{et al.}}}},\ }\bibfield  {title} {\bibinfo {title} {A
  gravitational wave observatory operating beyond the quantum shot-noise
  limit},\ }\href {https://doi.org/10.1038/nphys2083} {\bibfield  {journal}
  {\bibinfo  {journal} {Nature Physics}\ }\textbf {\bibinfo {volume} {7}},\
  \bibinfo {pages} {962} (\bibinfo {year} {2011})}\BibitemShut {NoStop}%
\bibitem [{\citenamefont {{Acernese \emph{et al.}}}(2019)}]{Acernese2019}%
  \BibitemOpen
  \bibfield  {author} {\bibinfo {author} {\bibfnamefont {F.}~\bibnamefont
  {{Acernese \emph{et al.}}}},\ }\bibfield  {title} {\bibinfo {title}
  {Increasing the astrophysical reach of the advanced virgo detector via the
  application of squeezed vacuum states of light},\ }\href
  {https://doi.org/10.1103/PhysRevLett.123.231108} {\bibfield  {journal}
  {\bibinfo  {journal} {Physical Review Letters}\ }\textbf {\bibinfo {volume}
  {123}},\ \bibinfo {pages} {231108} (\bibinfo {year} {2019})}\BibitemShut
  {NoStop}%
\bibitem [{\citenamefont {{Tse \emph{et al.}}}(2019)}]{Tse2019}%
  \BibitemOpen
  \bibfield  {author} {\bibinfo {author} {\bibfnamefont {M.}~\bibnamefont {{Tse
  \emph{et al.}}}},\ }\bibfield  {title} {\bibinfo {title} {Quantum-enhanced
  advanced {{LIGO}} detectors in the era of gravitational-wave astronomy},\
  }\href {https://doi.org/10.1103/PhysRevLett.123.231107} {\bibfield  {journal}
  {\bibinfo  {journal} {Physical Review Letters}\ }\textbf {\bibinfo {volume}
  {123}},\ \bibinfo {pages} {231107} (\bibinfo {year} {2019})}\BibitemShut
  {NoStop}%
\bibitem [{\citenamefont {{Yu \emph{et al.}}}(2020)}]{Yu2020}%
  \BibitemOpen
  \bibfield  {author} {\bibinfo {author} {\bibfnamefont {H.}~\bibnamefont {{Yu
  \emph{et al.}}}},\ }\bibfield  {title} {\bibinfo {title} {Quantum
  correlations between light and the kilogram-mass mirrors of {{LIGO}}},\
  }\href {https://doi.org/10.1038/s41586-020-2420-8} {\bibfield  {journal}
  {\bibinfo  {journal} {Nature}\ }\textbf {\bibinfo {volume} {583}},\ \bibinfo
  {pages} {43} (\bibinfo {year} {2020})}\BibitemShut {NoStop}%
\bibitem [{\citenamefont {Lough}\ \emph {et~al.}(2021)\citenamefont {Lough},
  \citenamefont {Schreiber}, \citenamefont {Bergamin}, \citenamefont {Grote},
  \citenamefont {Mehmet}, \citenamefont {Vahlbruch}, \citenamefont {Affeldt},
  \citenamefont {Brinkmann}, \citenamefont {Bisht}, \citenamefont {Kringel},
  \citenamefont {L{\"u}ck}, \citenamefont {Mukund}, \citenamefont {Nadji},
  \citenamefont {Sorazu}, \citenamefont {Strain}, \citenamefont {Weinert},\
  and\ \citenamefont {Danzmann}}]{Lough2021}%
  \BibitemOpen
  \bibfield  {author} {\bibinfo {author} {\bibfnamefont {J.}~\bibnamefont
  {Lough}}, \bibinfo {author} {\bibfnamefont {E.}~\bibnamefont {Schreiber}},
  \bibinfo {author} {\bibfnamefont {F.}~\bibnamefont {Bergamin}}, \bibinfo
  {author} {\bibfnamefont {H.}~\bibnamefont {Grote}}, \bibinfo {author}
  {\bibfnamefont {M.}~\bibnamefont {Mehmet}}, \bibinfo {author} {\bibfnamefont
  {H.}~\bibnamefont {Vahlbruch}}, \bibinfo {author} {\bibfnamefont
  {C.}~\bibnamefont {Affeldt}}, \bibinfo {author} {\bibfnamefont
  {M.}~\bibnamefont {Brinkmann}}, \bibinfo {author} {\bibfnamefont
  {A.}~\bibnamefont {Bisht}}, \bibinfo {author} {\bibfnamefont
  {V.}~\bibnamefont {Kringel}}, \bibinfo {author} {\bibfnamefont
  {H.}~\bibnamefont {L{\"u}ck}}, \bibinfo {author} {\bibfnamefont
  {N.}~\bibnamefont {Mukund}}, \bibinfo {author} {\bibfnamefont
  {S.}~\bibnamefont {Nadji}}, \bibinfo {author} {\bibfnamefont
  {B.}~\bibnamefont {Sorazu}}, \bibinfo {author} {\bibfnamefont
  {K.}~\bibnamefont {Strain}}, \bibinfo {author} {\bibfnamefont
  {M.}~\bibnamefont {Weinert}},\ and\ \bibinfo {author} {\bibfnamefont
  {K.}~\bibnamefont {Danzmann}},\ }\bibfield  {title} {\bibinfo {title} {First
  demonstration of 6 {{dB}} quantum noise reduction in a kilometer scale
  gravitational wave observatory},\ }\href
  {https://doi.org/10.1103/PhysRevLett.126.041102} {\bibfield  {journal}
  {\bibinfo  {journal} {Physical Review Letters}\ }\textbf {\bibinfo {volume}
  {126}},\ \bibinfo {pages} {041102} (\bibinfo {year} {2021})}\BibitemShut
  {NoStop}%
\bibitem [{\citenamefont {Yap}\ \emph {et~al.}(2020)\citenamefont {Yap},
  \citenamefont {Cripe}, \citenamefont {Mansell}, \citenamefont {McRae},
  \citenamefont {Ward}, \citenamefont {Slagmolen}, \citenamefont {Heu},
  \citenamefont {Follman}, \citenamefont {Cole}, \citenamefont {Corbitt},\ and\
  \citenamefont {McClelland}}]{Yap2020}%
  \BibitemOpen
  \bibfield  {author} {\bibinfo {author} {\bibfnamefont {M.~J.}\ \bibnamefont
  {Yap}}, \bibinfo {author} {\bibfnamefont {J.}~\bibnamefont {Cripe}}, \bibinfo
  {author} {\bibfnamefont {G.~L.}\ \bibnamefont {Mansell}}, \bibinfo {author}
  {\bibfnamefont {T.~G.}\ \bibnamefont {McRae}}, \bibinfo {author}
  {\bibfnamefont {R.~L.}\ \bibnamefont {Ward}}, \bibinfo {author}
  {\bibfnamefont {B.~J.~J.}\ \bibnamefont {Slagmolen}}, \bibinfo {author}
  {\bibfnamefont {P.}~\bibnamefont {Heu}}, \bibinfo {author} {\bibfnamefont
  {D.}~\bibnamefont {Follman}}, \bibinfo {author} {\bibfnamefont {G.~D.}\
  \bibnamefont {Cole}}, \bibinfo {author} {\bibfnamefont {T.}~\bibnamefont
  {Corbitt}},\ and\ \bibinfo {author} {\bibfnamefont {D.~E.}\ \bibnamefont
  {McClelland}},\ }\bibfield  {title} {\bibinfo {title} {Broadband reduction of
  quantum radiation pressure noise via squeezed light injection},\ }\href
  {https://doi.org/10.1038/s41566-019-0527-y} {\bibfield  {journal} {\bibinfo
  {journal} {Nature Photonics}\ }\textbf {\bibinfo {volume} {14}},\ \bibinfo
  {pages} {19} (\bibinfo {year} {2020})}\BibitemShut {NoStop}%
\bibitem [{\citenamefont {Backes}\ \emph {et~al.}(2021)\citenamefont {Backes},
  \citenamefont {Palken}, \citenamefont {Kenany}, \citenamefont {Brubaker},
  \citenamefont {Cahn}, \citenamefont {Droster}, \citenamefont {Hilton},
  \citenamefont {Ghosh}, \citenamefont {Jackson}, \citenamefont {Lamoreaux},
  \citenamefont {Leder}, \citenamefont {Lehnert}, \citenamefont {Lewis},
  \citenamefont {Malnou}, \citenamefont {Maruyama}, \citenamefont {Rapidis},
  \citenamefont {Simanovskaia}, \citenamefont {Singh}, \citenamefont {Speller},
  \citenamefont {Urdinaran}, \citenamefont {Vale}, \citenamefont {{van
  Assendelft}}, \citenamefont {{van Bibber}},\ and\ \citenamefont
  {Wang}}]{backesQuantumEnhancedSearch2021}%
  \BibitemOpen
  \bibfield  {author} {\bibinfo {author} {\bibfnamefont {K.~M.}\ \bibnamefont
  {Backes}}, \bibinfo {author} {\bibfnamefont {D.~A.}\ \bibnamefont {Palken}},
  \bibinfo {author} {\bibfnamefont {S.~A.}\ \bibnamefont {Kenany}}, \bibinfo
  {author} {\bibfnamefont {B.~M.}\ \bibnamefont {Brubaker}}, \bibinfo {author}
  {\bibfnamefont {S.~B.}\ \bibnamefont {Cahn}}, \bibinfo {author}
  {\bibfnamefont {A.}~\bibnamefont {Droster}}, \bibinfo {author} {\bibfnamefont
  {G.~C.}\ \bibnamefont {Hilton}}, \bibinfo {author} {\bibfnamefont
  {S.}~\bibnamefont {Ghosh}}, \bibinfo {author} {\bibfnamefont
  {H.}~\bibnamefont {Jackson}}, \bibinfo {author} {\bibfnamefont {S.~K.}\
  \bibnamefont {Lamoreaux}}, \bibinfo {author} {\bibfnamefont {A.~F.}\
  \bibnamefont {Leder}}, \bibinfo {author} {\bibfnamefont {K.~W.}\ \bibnamefont
  {Lehnert}}, \bibinfo {author} {\bibfnamefont {S.~M.}\ \bibnamefont {Lewis}},
  \bibinfo {author} {\bibfnamefont {M.}~\bibnamefont {Malnou}}, \bibinfo
  {author} {\bibfnamefont {R.~H.}\ \bibnamefont {Maruyama}}, \bibinfo {author}
  {\bibfnamefont {N.~M.}\ \bibnamefont {Rapidis}}, \bibinfo {author}
  {\bibfnamefont {M.}~\bibnamefont {Simanovskaia}}, \bibinfo {author}
  {\bibfnamefont {S.}~\bibnamefont {Singh}}, \bibinfo {author} {\bibfnamefont
  {D.~H.}\ \bibnamefont {Speller}}, \bibinfo {author} {\bibfnamefont
  {I.}~\bibnamefont {Urdinaran}}, \bibinfo {author} {\bibfnamefont {L.~R.}\
  \bibnamefont {Vale}}, \bibinfo {author} {\bibfnamefont {E.~C.}\ \bibnamefont
  {{van Assendelft}}}, \bibinfo {author} {\bibfnamefont {K.}~\bibnamefont {{van
  Bibber}}},\ and\ \bibinfo {author} {\bibfnamefont {H.}~\bibnamefont {Wang}},\
  }\bibfield  {title} {\bibinfo {title} {A quantum enhanced search for dark
  matter axions},\ }\href {https://doi.org/10.1038/s41586-021-03226-7}
  {\bibfield  {journal} {\bibinfo  {journal} {Nature}\ }\textbf {\bibinfo
  {volume} {590}},\ \bibinfo {pages} {238} (\bibinfo {year}
  {2021})}\BibitemShut {NoStop}%
\bibitem [{\citenamefont {Carney}\ \emph {et~al.}(2021)\citenamefont {Carney},
  \citenamefont {Krnjaic}, \citenamefont {Moore}, \citenamefont {Regal},
  \citenamefont {Afek}, \citenamefont {Bhave}, \citenamefont {Brubaker},
  \citenamefont {Corbitt}, \citenamefont {Cripe}, \citenamefont {Crisosto},
  \citenamefont {Geraci}, \citenamefont {Ghosh}, \citenamefont {Harris},
  \citenamefont {Hook}, \citenamefont {Kolb}, \citenamefont {Kunjummen},
  \citenamefont {Lang}, \citenamefont {Li}, \citenamefont {Lin}, \citenamefont
  {Liu}, \citenamefont {Lykken}, \citenamefont {Magrini}, \citenamefont
  {Manley}, \citenamefont {Matsumoto}, \citenamefont {Monte}, \citenamefont
  {Monteiro}, \citenamefont {Purdy}, \citenamefont {Riedel}, \citenamefont
  {Singh}, \citenamefont {Singh}, \citenamefont {Sinha}, \citenamefont
  {Taylor}, \citenamefont {Qin}, \citenamefont {Wilson},\ and\ \citenamefont
  {Zhao}}]{Carney2021}%
  \BibitemOpen
  \bibfield  {author} {\bibinfo {author} {\bibfnamefont {D.}~\bibnamefont
  {Carney}}, \bibinfo {author} {\bibfnamefont {G.}~\bibnamefont {Krnjaic}},
  \bibinfo {author} {\bibfnamefont {D.~C.}\ \bibnamefont {Moore}}, \bibinfo
  {author} {\bibfnamefont {C.~A.}\ \bibnamefont {Regal}}, \bibinfo {author}
  {\bibfnamefont {G.}~\bibnamefont {Afek}}, \bibinfo {author} {\bibfnamefont
  {S.}~\bibnamefont {Bhave}}, \bibinfo {author} {\bibfnamefont
  {B.}~\bibnamefont {Brubaker}}, \bibinfo {author} {\bibfnamefont
  {T.}~\bibnamefont {Corbitt}}, \bibinfo {author} {\bibfnamefont
  {J.}~\bibnamefont {Cripe}}, \bibinfo {author} {\bibfnamefont
  {N.}~\bibnamefont {Crisosto}}, \bibinfo {author} {\bibfnamefont
  {A.}~\bibnamefont {Geraci}}, \bibinfo {author} {\bibfnamefont
  {S.}~\bibnamefont {Ghosh}}, \bibinfo {author} {\bibfnamefont {J.~G.~E.}\
  \bibnamefont {Harris}}, \bibinfo {author} {\bibfnamefont {A.}~\bibnamefont
  {Hook}}, \bibinfo {author} {\bibfnamefont {E.~W.}\ \bibnamefont {Kolb}},
  \bibinfo {author} {\bibfnamefont {J.}~\bibnamefont {Kunjummen}}, \bibinfo
  {author} {\bibfnamefont {R.~F.}\ \bibnamefont {Lang}}, \bibinfo {author}
  {\bibfnamefont {T.}~\bibnamefont {Li}}, \bibinfo {author} {\bibfnamefont
  {T.}~\bibnamefont {Lin}}, \bibinfo {author} {\bibfnamefont {Z.}~\bibnamefont
  {Liu}}, \bibinfo {author} {\bibfnamefont {J.}~\bibnamefont {Lykken}},
  \bibinfo {author} {\bibfnamefont {L.}~\bibnamefont {Magrini}}, \bibinfo
  {author} {\bibfnamefont {J.}~\bibnamefont {Manley}}, \bibinfo {author}
  {\bibfnamefont {N.}~\bibnamefont {Matsumoto}}, \bibinfo {author}
  {\bibfnamefont {A.}~\bibnamefont {Monte}}, \bibinfo {author} {\bibfnamefont
  {F.}~\bibnamefont {Monteiro}}, \bibinfo {author} {\bibfnamefont
  {T.}~\bibnamefont {Purdy}}, \bibinfo {author} {\bibfnamefont {C.~J.}\
  \bibnamefont {Riedel}}, \bibinfo {author} {\bibfnamefont {R.}~\bibnamefont
  {Singh}}, \bibinfo {author} {\bibfnamefont {S.}~\bibnamefont {Singh}},
  \bibinfo {author} {\bibfnamefont {K.}~\bibnamefont {Sinha}}, \bibinfo
  {author} {\bibfnamefont {J.~M.}\ \bibnamefont {Taylor}}, \bibinfo {author}
  {\bibfnamefont {J.}~\bibnamefont {Qin}}, \bibinfo {author} {\bibfnamefont
  {D.~J.}\ \bibnamefont {Wilson}},\ and\ \bibinfo {author} {\bibfnamefont
  {Y.}~\bibnamefont {Zhao}},\ }\bibfield  {title} {\bibinfo {title} {Mechanical
  quantum sensing in the search for dark matter},\ }\href
  {https://doi.org/10.1088/2058-9565/abcfcd} {\bibfield  {journal} {\bibinfo
  {journal} {Quantum Science and Technology}\ }\textbf {\bibinfo {volume}
  {6}},\ \bibinfo {pages} {024002} (\bibinfo {year} {2021})}\BibitemShut
  {NoStop}%
\bibitem [{\citenamefont {Taylor}\ \emph {et~al.}(2013)\citenamefont {Taylor},
  \citenamefont {Janousek}, \citenamefont {Daria}, \citenamefont {Knittel},
  \citenamefont {Hage}, \citenamefont {Bachor},\ and\ \citenamefont
  {Bowen}}]{Taylor2013}%
  \BibitemOpen
  \bibfield  {author} {\bibinfo {author} {\bibfnamefont {M.~A.}\ \bibnamefont
  {Taylor}}, \bibinfo {author} {\bibfnamefont {J.}~\bibnamefont {Janousek}},
  \bibinfo {author} {\bibfnamefont {V.}~\bibnamefont {Daria}}, \bibinfo
  {author} {\bibfnamefont {J.}~\bibnamefont {Knittel}}, \bibinfo {author}
  {\bibfnamefont {B.}~\bibnamefont {Hage}}, \bibinfo {author} {\bibfnamefont
  {H.-A.}\ \bibnamefont {Bachor}},\ and\ \bibinfo {author} {\bibfnamefont
  {W.~P.}\ \bibnamefont {Bowen}},\ }\bibfield  {title} {\bibinfo {title}
  {Biological measurement beyond the quantum limit},\ }\href
  {https://doi.org/10.1038/nphoton.2012.346} {\bibfield  {journal} {\bibinfo
  {journal} {Nature Photonics}\ }\textbf {\bibinfo {volume} {7}},\ \bibinfo
  {pages} {229} (\bibinfo {year} {2013})}\BibitemShut {NoStop}%
\bibitem [{\citenamefont {Taylor}\ and\ \citenamefont
  {Bowen}(2016)}]{Taylor2016}%
  \BibitemOpen
  \bibfield  {author} {\bibinfo {author} {\bibfnamefont {M.~A.}\ \bibnamefont
  {Taylor}}\ and\ \bibinfo {author} {\bibfnamefont {W.~P.}\ \bibnamefont
  {Bowen}},\ }\bibfield  {title} {\bibinfo {title} {Quantum metrology and its
  application in biology},\ }\href
  {https://doi.org/10.1016/j.physrep.2015.12.002} {\bibfield  {journal}
  {\bibinfo  {journal} {Physics Reports}\ }\textbf {\bibinfo {volume} {615}},\
  \bibinfo {pages} {1} (\bibinfo {year} {2016})},\ \Eprint
  {https://arxiv.org/abs/1409.0950} {arXiv:1409.0950} \BibitemShut {NoStop}%
\bibitem [{\citenamefont {Li}\ \emph {et~al.}(2018)\citenamefont {Li},
  \citenamefont {B{\'i}lek}, \citenamefont {Hoff}, \citenamefont {Madsen},
  \citenamefont {Forstner}, \citenamefont {Prakash}, \citenamefont
  {Sch{\"a}fermeier}, \citenamefont {Gehring}, \citenamefont {Bowen},\ and\
  \citenamefont {Andersen}}]{Li2018}%
  \BibitemOpen
  \bibfield  {author} {\bibinfo {author} {\bibfnamefont {B.-B.}\ \bibnamefont
  {Li}}, \bibinfo {author} {\bibfnamefont {J.}~\bibnamefont {B{\'i}lek}},
  \bibinfo {author} {\bibfnamefont {U.~B.}\ \bibnamefont {Hoff}}, \bibinfo
  {author} {\bibfnamefont {L.~S.}\ \bibnamefont {Madsen}}, \bibinfo {author}
  {\bibfnamefont {S.}~\bibnamefont {Forstner}}, \bibinfo {author}
  {\bibfnamefont {V.}~\bibnamefont {Prakash}}, \bibinfo {author} {\bibfnamefont
  {C.}~\bibnamefont {Sch{\"a}fermeier}}, \bibinfo {author} {\bibfnamefont
  {T.}~\bibnamefont {Gehring}}, \bibinfo {author} {\bibfnamefont {W.~P.}\
  \bibnamefont {Bowen}},\ and\ \bibinfo {author} {\bibfnamefont {U.~L.}\
  \bibnamefont {Andersen}},\ }\bibfield  {title} {\bibinfo {title} {Quantum
  enhanced optomechanical magnetometry},\ }\href
  {https://doi.org/10.1364/OPTICA.5.000850} {\bibfield  {journal} {\bibinfo
  {journal} {Optica}\ }\textbf {\bibinfo {volume} {5}},\ \bibinfo {pages} {850}
  (\bibinfo {year} {2018})},\ \Eprint {https://arxiv.org/abs/1802.09738}
  {arXiv:1802.09738} \BibitemShut {NoStop}%
\bibitem [{\citenamefont {Tsang}\ \emph {et~al.}(2016)\citenamefont {Tsang},
  \citenamefont {Nair},\ and\ \citenamefont
  {Lu}}]{tsangQuantumTheorySuperresolution2016}%
  \BibitemOpen
  \bibfield  {author} {\bibinfo {author} {\bibfnamefont {M.}~\bibnamefont
  {Tsang}}, \bibinfo {author} {\bibfnamefont {R.}~\bibnamefont {Nair}},\ and\
  \bibinfo {author} {\bibfnamefont {X.-M.}\ \bibnamefont {Lu}},\ }\bibfield
  {title} {\bibinfo {title} {Quantum theory of superresolution for two
  incoherent optical point sources},\ }\href@noop {} {\bibfield  {journal}
  {\bibinfo  {journal} {Physical Review X}\ }\textbf {\bibinfo {volume} {6}},\
  \bibinfo {pages} {031033} (\bibinfo {year} {2016})}\BibitemShut {NoStop}%
\bibitem [{\citenamefont {Bentley}\ \emph {et~al.}(2023)\citenamefont
  {Bentley}, \citenamefont {Nurdin}, \citenamefont {Chen}, \citenamefont {Li},\
  and\ \citenamefont {Miao}}]{bentleyDesigningOptimalLinear2023}%
  \BibitemOpen
  \bibfield  {author} {\bibinfo {author} {\bibfnamefont {J.}~\bibnamefont
  {Bentley}}, \bibinfo {author} {\bibfnamefont {H.}~\bibnamefont {Nurdin}},
  \bibinfo {author} {\bibfnamefont {Y.}~\bibnamefont {Chen}}, \bibinfo {author}
  {\bibfnamefont {X.}~\bibnamefont {Li}},\ and\ \bibinfo {author}
  {\bibfnamefont {H.}~\bibnamefont {Miao}},\ }\bibfield  {title} {\bibinfo
  {title} {Designing {{Optimal Linear Detectors}}: {{A Bottom-Up Approach}}},\
  }\href {https://doi.org/10.1103/PhysRevApplied.19.034009} {\bibfield
  {journal} {\bibinfo  {journal} {Physical Review Applied}\ }\textbf {\bibinfo
  {volume} {19}},\ \bibinfo {pages} {034009} (\bibinfo {year}
  {2023})}\BibitemShut {NoStop}%
\bibitem [{\citenamefont {Braginsky}\ \emph {et~al.}(2000)\citenamefont
  {Braginsky}, \citenamefont {Gorodetsky}, \citenamefont {Thorne},\ and\
  \citenamefont {Khalili}}]{Braginsky2000}%
  \BibitemOpen
  \bibfield  {author} {\bibinfo {author} {\bibfnamefont {V.~B.}\ \bibnamefont
  {Braginsky}}, \bibinfo {author} {\bibfnamefont {M.~L.}\ \bibnamefont
  {Gorodetsky}}, \bibinfo {author} {\bibfnamefont {K.~S.}\ \bibnamefont
  {Thorne}},\ and\ \bibinfo {author} {\bibfnamefont {F.~Y.}\ \bibnamefont
  {Khalili}},\ }\bibfield  {title} {\bibinfo {title} {Energetic {{Quantum
  Limit}} in {{Large-Scale Interferometers}}},\ }in\ \href@noop {} {\emph
  {\bibinfo {booktitle} {Gravitational Waves. Third Edoardo Amaldi Conference,
  Pasadena, California, 12-16 {{July}}}}},\ \bibinfo {editor} {edited by\
  \bibinfo {editor} {\bibnamefont {{S.Meshkov}}}}\ (\bibinfo  {publisher}
  {{Melville NY:AIP Conf. Proc. 523}},\ \bibinfo {year} {2000})\ pp.\ \bibinfo
  {pages} {180--189}\BibitemShut {NoStop}%
\bibitem [{\citenamefont {Tsang}\ \emph {et~al.}(2011)\citenamefont {Tsang},
  \citenamefont {Wiseman},\ and\ \citenamefont {Caves}}]{Tsang2011}%
  \BibitemOpen
  \bibfield  {author} {\bibinfo {author} {\bibfnamefont {M.}~\bibnamefont
  {Tsang}}, \bibinfo {author} {\bibfnamefont {H.~M.}\ \bibnamefont {Wiseman}},\
  and\ \bibinfo {author} {\bibfnamefont {C.~M.}\ \bibnamefont {Caves}},\
  }\bibfield  {title} {\bibinfo {title} {Fundamental {{Quantum Limit}} to
  {{Waveform Estimation}}},\ }\href
  {https://doi.org/10.1103/PhysRevLett.106.090401} {\bibfield  {journal}
  {\bibinfo  {journal} {Physical Review Letters}\ }\textbf {\bibinfo {volume}
  {106}},\ \bibinfo {pages} {090401} (\bibinfo {year} {2011})},\ \Eprint
  {https://arxiv.org/abs/1006.5407} {arXiv:1006.5407} \BibitemShut {NoStop}%
\bibitem [{\citenamefont {Miao}\ \emph {et~al.}(2017)\citenamefont {Miao},
  \citenamefont {Adhikari}, \citenamefont {Ma}, \citenamefont {Pang},\ and\
  \citenamefont {Chen}}]{Miao2017}%
  \BibitemOpen
  \bibfield  {author} {\bibinfo {author} {\bibfnamefont {H.}~\bibnamefont
  {Miao}}, \bibinfo {author} {\bibfnamefont {R.~X.}\ \bibnamefont {Adhikari}},
  \bibinfo {author} {\bibfnamefont {Y.}~\bibnamefont {Ma}}, \bibinfo {author}
  {\bibfnamefont {B.}~\bibnamefont {Pang}},\ and\ \bibinfo {author}
  {\bibfnamefont {Y.}~\bibnamefont {Chen}},\ }\bibfield  {title} {\bibinfo
  {title} {Towards the {{Fundamental Quantum Limit}} of {{Linear Measurements}}
  of {{Classical Signals}}},\ }\href
  {https://doi.org/10.1103/PhysRevLett.119.050801} {\bibfield  {journal}
  {\bibinfo  {journal} {Physical Review Letters}\ }\textbf {\bibinfo {volume}
  {119}},\ \bibinfo {pages} {050801} (\bibinfo {year} {2017})},\ \Eprint
  {https://arxiv.org/abs/1608.00766} {arXiv:1608.00766} \BibitemShut {NoStop}%
\bibitem [{\citenamefont {Dorner}\ \emph {et~al.}(2009)\citenamefont {Dorner},
  \citenamefont {Demkowicz-Dobrza{\'{n}}ski}, \citenamefont {Smith},
  \citenamefont {Lundeen}, \citenamefont {Wasilewski}, \citenamefont
  {Banaszek},\ and\ \citenamefont {Walmsley}}]{Dorner2009}%
  \BibitemOpen
  \bibfield  {author} {\bibinfo {author} {\bibfnamefont {U.}~\bibnamefont
  {Dorner}}, \bibinfo {author} {\bibfnamefont {R.}~\bibnamefont
  {Demkowicz-Dobrza{\'{n}}ski}}, \bibinfo {author} {\bibfnamefont {B.~J.}\
  \bibnamefont {Smith}}, \bibinfo {author} {\bibfnamefont {J.~S.}\ \bibnamefont
  {Lundeen}}, \bibinfo {author} {\bibfnamefont {W.}~\bibnamefont {Wasilewski}},
  \bibinfo {author} {\bibfnamefont {K.}~\bibnamefont {Banaszek}},\ and\
  \bibinfo {author} {\bibfnamefont {I.~A.}\ \bibnamefont {Walmsley}},\
  }\bibfield  {title} {\bibinfo {title} {{Optimal Quantum Phase Estimation}},\
  }\href {https://doi.org/10.1103/PhysRevLett.102.040403} {\bibfield  {journal}
  {\bibinfo  {journal} {Physical Review Letters}\ }\textbf {\bibinfo {volume}
  {102}},\ \bibinfo {pages} {040403} (\bibinfo {year} {2009})},\ \Eprint
  {https://arxiv.org/abs/0807.3659} {arXiv:0807.3659} \BibitemShut {NoStop}%
\bibitem [{\citenamefont {{Demkowicz-Dobrza{\'n}ski}}\ \emph
  {et~al.}(2013)\citenamefont {{Demkowicz-Dobrza{\'n}ski}}, \citenamefont
  {Banaszek},\ and\ \citenamefont
  {\mbox{Schnabel}}}]{Demkowicz-Dobrzanski2013}%
  \BibitemOpen
  \bibfield  {author} {\bibinfo {author} {\bibfnamefont {R.}~\bibnamefont
  {{Demkowicz-Dobrza{\'n}ski}}}, \bibinfo {author} {\bibfnamefont
  {K.}~\bibnamefont {Banaszek}},\ and\ \bibinfo {author} {\bibfnamefont
  {R.}~\bibnamefont {\mbox{Schnabel}}},\ }\bibfield  {title} {\bibinfo {title}
  {Fundamental quantum interferometry bound for the squeezed-light-enhanced
  gravitational wave detector {{GEO}} 600},\ }\href
  {https://doi.org/10.1103/PhysRevA.88.041802} {\bibfield  {journal} {\bibinfo
  {journal} {Physical Review A}\ }\textbf {\bibinfo {volume} {88}},\ \bibinfo
  {pages} {041802(R)} (\bibinfo {year} {2013})}\BibitemShut {NoStop}%
\bibitem [{\citenamefont {{Demkowicz-Dobrza{\'n}ski}}\ \emph
  {et~al.}(2015)\citenamefont {{Demkowicz-Dobrza{\'n}ski}}, \citenamefont
  {Jarzyna},\ and\ \citenamefont {Ko{\l}odynski}}]{Demkowicz-Dobrzanski2015}%
  \BibitemOpen
  \bibfield  {author} {\bibinfo {author} {\bibfnamefont {R.}~\bibnamefont
  {{Demkowicz-Dobrza{\'n}ski}}}, \bibinfo {author} {\bibfnamefont
  {M.}~\bibnamefont {Jarzyna}},\ and\ \bibinfo {author} {\bibfnamefont
  {J.}~\bibnamefont {Ko{\l}odynski}},\ }\bibfield  {title} {\bibinfo {title}
  {Quantum {{Limits}} in {{Optical Interferometry}}},\ }\href@noop {}
  {\bibfield  {journal} {\bibinfo  {journal} {Progress in Optics}\ }\textbf
  {\bibinfo {volume} {60}},\ \bibinfo {pages} {345} (\bibinfo {year}
  {2015})}\BibitemShut {NoStop}%
\bibitem [{\citenamefont {Miao}\ \emph {et~al.}(2019)\citenamefont {Miao},
  \citenamefont {Smith},\ and\ \citenamefont {Evans}}]{Miao2019}%
  \BibitemOpen
  \bibfield  {author} {\bibinfo {author} {\bibfnamefont {H.}~\bibnamefont
  {Miao}}, \bibinfo {author} {\bibfnamefont {N.~D.}\ \bibnamefont {Smith}},\
  and\ \bibinfo {author} {\bibfnamefont {M.}~\bibnamefont {Evans}},\ }\bibfield
   {title} {\bibinfo {title} {Quantum {{Limit}} for {{Laser Interferometric
  Gravitational-Wave Detectors}} from {{Optical Dissipation}}},\ }\href
  {https://doi.org/10.1103/PhysRevX.9.011053} {\bibfield  {journal} {\bibinfo
  {journal} {Physical Review X}\ }\textbf {\bibinfo {volume} {9}},\ \bibinfo
  {pages} {011053} (\bibinfo {year} {2019})}\BibitemShut {NoStop}%
\bibitem [{\citenamefont {Dowling}(2008)}]{Dowling2008}%
  \BibitemOpen
  \bibfield  {author} {\bibinfo {author} {\bibfnamefont {J.~P.}\ \bibnamefont
  {Dowling}},\ }\bibfield  {title} {\bibinfo {title} {{Quantum optical
  metrology -- the lowdown on high-N00N states}},\ }\href
  {https://doi.org/10.1080/00107510802091298} {\bibfield  {journal} {\bibinfo
  {journal} {Contemporary Physics}\ }\textbf {\bibinfo {volume} {49}},\
  \bibinfo {pages} {125} (\bibinfo {year} {2008})},\ \Eprint
  {https://arxiv.org/abs/0904.0163} {arXiv:0904.0163} \BibitemShut {NoStop}%
\bibitem [{\citenamefont {Rehbein}\ \emph {et~al.}(2005)\citenamefont
  {Rehbein}, \citenamefont {Harms}, \citenamefont {Schnabel},\ and\
  \citenamefont {Danzmann}}]{Rehbein2005}%
  \BibitemOpen
  \bibfield  {author} {\bibinfo {author} {\bibfnamefont {H.}~\bibnamefont
  {Rehbein}}, \bibinfo {author} {\bibfnamefont {J.}~\bibnamefont {Harms}},
  \bibinfo {author} {\bibfnamefont {R.}~\bibnamefont {Schnabel}},\ and\
  \bibinfo {author} {\bibfnamefont {K.}~\bibnamefont {Danzmann}},\ }\bibfield
  {title} {\bibinfo {title} {Optical {{Transfer Functions}} of {{Kerr Nonlinear
  Cavities}} and {{Interferometers}}},\ }\href
  {https://doi.org/10.1103/PhysRevLett.95.193001} {\bibfield  {journal}
  {\bibinfo  {journal} {Physical Review Letters}\ }\textbf {\bibinfo {volume}
  {95}},\ \bibinfo {pages} {193001} (\bibinfo {year} {2005})}\BibitemShut
  {NoStop}%
\bibitem [{\citenamefont {Peano}\ \emph {et~al.}(2015)\citenamefont {Peano},
  \citenamefont {Schwefel}, \citenamefont {Marquardt},\ and\ \citenamefont
  {Marquardt}}]{V.Peano2015}%
  \BibitemOpen
  \bibfield  {author} {\bibinfo {author} {\bibfnamefont {V.}~\bibnamefont
  {Peano}}, \bibinfo {author} {\bibfnamefont {H.~G.~L.}\ \bibnamefont
  {Schwefel}}, \bibinfo {author} {\bibfnamefont {C.}~\bibnamefont
  {Marquardt}},\ and\ \bibinfo {author} {\bibfnamefont {F.}~\bibnamefont
  {Marquardt}},\ }\bibfield  {title} {\bibinfo {title} {Intracavity {{Squeezing
  Can Enhance Quantum-Limited Optomechanical Position Detection}} through
  {{Deamplification}}},\ }\href
  {https://doi.org/10.1103/PhysRevLett.115.243603} {\bibfield  {journal}
  {\bibinfo  {journal} {Physical Review Letters}\ }\textbf {\bibinfo {volume}
  {115}},\ \bibinfo {pages} {243603} (\bibinfo {year} {2015})},\ \Eprint
  {https://arxiv.org/abs/1502.06423} {arXiv:1502.06423} \BibitemShut {NoStop}%
\bibitem [{\citenamefont {Korobko}\ \emph {et~al.}(2017)\citenamefont
  {Korobko}, \citenamefont {Kleybolte}, \citenamefont {Ast}, \citenamefont
  {Miao}, \citenamefont {Chen},\ and\ \citenamefont
  {Schnabel}}]{korobko2017beating}%
  \BibitemOpen
  \bibfield  {author} {\bibinfo {author} {\bibfnamefont {M.}~\bibnamefont
  {Korobko}}, \bibinfo {author} {\bibfnamefont {L.}~\bibnamefont {Kleybolte}},
  \bibinfo {author} {\bibfnamefont {S.}~\bibnamefont {Ast}}, \bibinfo {author}
  {\bibfnamefont {H.}~\bibnamefont {Miao}}, \bibinfo {author} {\bibfnamefont
  {Y.}~\bibnamefont {Chen}},\ and\ \bibinfo {author} {\bibfnamefont
  {R.}~\bibnamefont {Schnabel}},\ }\bibfield  {title} {\bibinfo {title}
  {Beating the {{Standard Sensitivity-Bandwidth Limit}} of {{Cavity-Enhanced
  Interferometers}} with {{Internal Squeezed-Light Generation}}},\ }\href
  {https://doi.org/10.1103/PhysRevLett.118.143601} {\bibfield  {journal}
  {\bibinfo  {journal} {Physical Review Letters}\ }\textbf {\bibinfo {volume}
  {118}},\ \bibinfo {pages} {143601} (\bibinfo {year} {2017})}\BibitemShut
  {NoStop}%
\bibitem [{\citenamefont {Korobko}\ \emph {et~al.}(2018)\citenamefont
  {Korobko}, \citenamefont {Khalili},\ and\ \citenamefont
  {Schnabel}}]{korobko2018engineering}%
  \BibitemOpen
  \bibfield  {author} {\bibinfo {author} {\bibfnamefont {M.}~\bibnamefont
  {Korobko}}, \bibinfo {author} {\bibfnamefont {F.~Y.}\ \bibnamefont
  {Khalili}},\ and\ \bibinfo {author} {\bibfnamefont {R.}~\bibnamefont
  {Schnabel}},\ }\bibfield  {title} {\bibinfo {title} {Engineering the optical
  spring via intra-cavity optical-parametric amplification},\ }\href
  {https://doi.org/10.1016/j.physleta.2017.08.008} {\bibfield  {journal}
  {\bibinfo  {journal} {Physics Letters A}\ }\textbf {\bibinfo {volume}
  {382}},\ \bibinfo {pages} {2238} (\bibinfo {year} {2018})},\ \Eprint
  {https://arxiv.org/abs/1709.03055} {arXiv:1709.03055} \BibitemShut {NoStop}%
\bibitem [{\citenamefont {Korobko}\ \emph {et~al.}(2019)\citenamefont
  {Korobko}, \citenamefont {Ma}, \citenamefont {Chen},\ and\ \citenamefont
  {Schnabel}}]{korobko2019quantum}%
  \BibitemOpen
  \bibfield  {author} {\bibinfo {author} {\bibfnamefont {M.}~\bibnamefont
  {Korobko}}, \bibinfo {author} {\bibfnamefont {Y.}~\bibnamefont {Ma}},
  \bibinfo {author} {\bibfnamefont {Y.}~\bibnamefont {Chen}},\ and\ \bibinfo
  {author} {\bibfnamefont {R.}~\bibnamefont {Schnabel}},\ }\bibfield  {title}
  {\bibinfo {title} {Quantum expander for gravitational-wave observatories},\
  }\href {https://doi.org/10.1038/s41377-019-0230-2} {\bibfield  {journal}
  {\bibinfo  {journal} {Light: Science \& Applications}\ }\textbf {\bibinfo
  {volume} {8}},\ \bibinfo {pages} {118} (\bibinfo {year} {2019})},\ \Eprint
  {https://arxiv.org/abs/1903.05930} {arXiv:1903.05930} \BibitemShut {NoStop}%
\bibitem [{\citenamefont {Adya}\ \emph {et~al.}(2020)\citenamefont {Adya},
  \citenamefont {Yap}, \citenamefont {T{\"o}yr{\"a}}, \citenamefont {McRae},
  \citenamefont {Altin}, \citenamefont {Sarre}, \citenamefont {Meijerink},
  \citenamefont {Kijbunchoo}, \citenamefont {Slagmolen}, \citenamefont {Ward},\
  and\ \citenamefont {McClelland}}]{adyaQuantumEnhancedKHz2020}%
  \BibitemOpen
  \bibfield  {author} {\bibinfo {author} {\bibfnamefont {V.~B.}\ \bibnamefont
  {Adya}}, \bibinfo {author} {\bibfnamefont {M.~J.}\ \bibnamefont {Yap}},
  \bibinfo {author} {\bibfnamefont {D.}~\bibnamefont {T{\"o}yr{\"a}}}, \bibinfo
  {author} {\bibfnamefont {T.~G.}\ \bibnamefont {McRae}}, \bibinfo {author}
  {\bibfnamefont {P.~A.}\ \bibnamefont {Altin}}, \bibinfo {author}
  {\bibfnamefont {L.~K.}\ \bibnamefont {Sarre}}, \bibinfo {author}
  {\bibfnamefont {M.}~\bibnamefont {Meijerink}}, \bibinfo {author}
  {\bibfnamefont {N.}~\bibnamefont {Kijbunchoo}}, \bibinfo {author}
  {\bibfnamefont {B.~J.~J.}\ \bibnamefont {Slagmolen}}, \bibinfo {author}
  {\bibfnamefont {R.~L.}\ \bibnamefont {Ward}},\ and\ \bibinfo {author}
  {\bibfnamefont {D.~E.}\ \bibnamefont {McClelland}},\ }\bibfield  {title}
  {\bibinfo {title} {Quantum enhanced {{kHz}} gravitational wave detector with
  internal squeezing},\ }\href {https://doi.org/10.1088/1361-6382/ab7615}
  {\bibfield  {journal} {\bibinfo  {journal} {Classical and Quantum Gravity}\
  }\textbf {\bibinfo {volume} {37}},\ \bibinfo {pages} {07LT02} (\bibinfo
  {year} {2020})}\BibitemShut {NoStop}%
\bibitem [{\citenamefont {Caves}\ and\ \citenamefont
  {Schumaker}(1985)}]{Caves1985a}%
  \BibitemOpen
  \bibfield  {author} {\bibinfo {author} {\bibfnamefont {C.~M.}\ \bibnamefont
  {Caves}}\ and\ \bibinfo {author} {\bibfnamefont {B.~L.}\ \bibnamefont
  {Schumaker}},\ }\bibfield  {title} {\bibinfo {title} {New formalism for
  two-photon quantum optics. {{I}}. {{Quadrature}} phases and squeezed
  states},\ }\href {https://doi.org/10.1103/PhysRevA.31.3068} {\bibfield
  {journal} {\bibinfo  {journal} {Physical Review A}\ }\textbf {\bibinfo
  {volume} {31}},\ \bibinfo {pages} {3068} (\bibinfo {year}
  {1985})}\BibitemShut {NoStop}%
\bibitem [{\citenamefont {Danilishin}\ and\ \citenamefont
  {Khalili}(2012)}]{Danilishin2012}%
  \BibitemOpen
  \bibfield  {author} {\bibinfo {author} {\bibfnamefont {S.~L.}\ \bibnamefont
  {Danilishin}}\ and\ \bibinfo {author} {\bibfnamefont {F.~Y.}\ \bibnamefont
  {Khalili}},\ }\bibfield  {title} {\bibinfo {title} {Quantum {{Measurement
  Theory}} in {{Gravitational-Wave Detectors}}},\ }\bibfield  {journal}
  {\bibinfo  {journal} {Living Reviews in Relativity}\ }\textbf {\bibinfo
  {volume} {15}},\ \href {https://doi.org/10.12942/lrr-2012-5}
  {10.12942/lrr-2012-5} (\bibinfo {year} {2012})\BibitemShut {NoStop}%
\bibitem [{\citenamefont {Schumaker}\ and\ \citenamefont
  {Caves}(1985)}]{Schumaker1985a}%
  \BibitemOpen
  \bibfield  {author} {\bibinfo {author} {\bibfnamefont {B.~L.}\ \bibnamefont
  {Schumaker}}\ and\ \bibinfo {author} {\bibfnamefont {C.~M.}\ \bibnamefont
  {Caves}},\ }\bibfield  {title} {\bibinfo {title} {New formalism for
  two-photon quantum optics. {{II}}. {{Mathematical}} foundation and compact
  notation},\ }\href {https://doi.org/10.1103/PhysRevA.31.3093} {\bibfield
  {journal} {\bibinfo  {journal} {Physical Review A}\ }\textbf {\bibinfo
  {volume} {31}},\ \bibinfo {pages} {3093} (\bibinfo {year}
  {1985})}\BibitemShut {NoStop}%
\bibitem [{\citenamefont {Stoler}(1970)}]{Stoler1970}%
  \BibitemOpen
  \bibfield  {author} {\bibinfo {author} {\bibfnamefont {D.}~\bibnamefont
  {Stoler}},\ }\bibfield  {title} {\bibinfo {title} {{Equivalence Classes of
  Minimum Uncertainty Packets}},\ }\href
  {https://doi.org/10.1103/PhysRevD.1.3217} {\bibfield  {journal} {\bibinfo
  {journal} {Physical Review D}\ }\textbf {\bibinfo {volume} {1}},\ \bibinfo
  {pages} {3217} (\bibinfo {year} {1970})}\BibitemShut {NoStop}%
\bibitem [{\citenamefont {Korobko}\ \emph {et~al.}(2023)\citenamefont
  {Korobko}, \citenamefont {S\"udbeck}, \citenamefont {Steinlechner},\ and\
  \citenamefont {Schnabel}}]{korobko2023long}%
  \BibitemOpen
  \bibfield  {author} {\bibinfo {author} {\bibfnamefont {M.}~\bibnamefont
  {Korobko}}, \bibinfo {author} {\bibfnamefont {J.}~\bibnamefont {S\"udbeck}},
  \bibinfo {author} {\bibfnamefont {S.}~\bibnamefont {Steinlechner}},\ and\
  \bibinfo {author} {\bibfnamefont {R.}~\bibnamefont {Schnabel}},\ }\bibfield
  {title} {\bibinfo {title} {Fundamental sensitivity limit in cavity-enhanced
  interferometers with external and internal squeezing},\ }\href@noop {}
  {\bibfield  {journal} {\bibinfo  {journal} {to be published}\ } (\bibinfo
  {year} {2023})}\BibitemShut {NoStop}%
\bibitem [{\citenamefont {Kimble}\ \emph {et~al.}(2001)\citenamefont {Kimble},
  \citenamefont {Levin}, \citenamefont {Matsko}, \citenamefont {Thorne},\ and\
  \citenamefont {Vyatchanin}}]{Kimble2001}%
  \BibitemOpen
  \bibfield  {author} {\bibinfo {author} {\bibfnamefont {H.~J.}\ \bibnamefont
  {Kimble}}, \bibinfo {author} {\bibfnamefont {Y.}~\bibnamefont {Levin}},
  \bibinfo {author} {\bibfnamefont {A.~B.}\ \bibnamefont {Matsko}}, \bibinfo
  {author} {\bibfnamefont {K.~S.}\ \bibnamefont {Thorne}},\ and\ \bibinfo
  {author} {\bibfnamefont {S.~P.}\ \bibnamefont {Vyatchanin}},\ }\bibfield
  {title} {\bibinfo {title} {{Conversion of conventional gravitational-wave
  interferometers into quantum nondemolition interferometers by modifying their
  input and/or output optics}},\ }\href
  {https://doi.org/10.1103/PhysRevD.65.022002} {\bibfield  {journal} {\bibinfo
  {journal} {Physical Review D}\ }\textbf {\bibinfo {volume} {65}},\ \bibinfo
  {pages} {022002} (\bibinfo {year} {2001})}\BibitemShut {NoStop}%
\bibitem [{\citenamefont {Milburn}\ and\ \citenamefont
  {Walls}(1981)}]{Milburn1981}%
  \BibitemOpen
  \bibfield  {author} {\bibinfo {author} {\bibfnamefont {G.}~\bibnamefont
  {Milburn}}\ and\ \bibinfo {author} {\bibfnamefont {D.}~\bibnamefont
  {Walls}},\ }\bibfield  {title} {\bibinfo {title} {Production of squeezed
  states in a degenerate parametric amplifier},\ }\href
  {https://doi.org/10.1016/0030-4018(81)90232-7} {\bibfield  {journal}
  {\bibinfo  {journal} {Optics Communications}\ }\textbf {\bibinfo {volume}
  {39}},\ \bibinfo {pages} {401} (\bibinfo {year} {1981})}\BibitemShut
  {NoStop}%
\bibitem [{\citenamefont {Collett}\ and\ \citenamefont
  {Gardiner}(1984)}]{Collett1984}%
  \BibitemOpen
  \bibfield  {author} {\bibinfo {author} {\bibfnamefont {M.~J.}\ \bibnamefont
  {Collett}}\ and\ \bibinfo {author} {\bibfnamefont {C.~W.}\ \bibnamefont
  {Gardiner}},\ }\bibfield  {title} {\bibinfo {title} {Squeezing of intracavity
  and traveling-wave light fields produced in parametric amplification},\
  }\href {https://doi.org/10.1103/PhysRevA.30.1386} {\bibfield  {journal}
  {\bibinfo  {journal} {Physical Review A}\ }\textbf {\bibinfo {volume} {30}},\
  \bibinfo {pages} {1386} (\bibinfo {year} {1984})}\BibitemShut {NoStop}%
\bibitem [{\citenamefont {Schnabel}(2017)}]{Schnabel2017}%
  \BibitemOpen
  \bibfield  {author} {\bibinfo {author} {\bibfnamefont {R.}~\bibnamefont
  {Schnabel}},\ }\bibfield  {title} {\bibinfo {title} {Squeezed states of light
  and their applications in laser interferometers},\ }\href
  {https://doi.org/10.1016/j.physrep.2017.04.001} {\bibfield  {journal}
  {\bibinfo  {journal} {Physics Reports}\ }\textbf {\bibinfo {volume} {684}},\
  \bibinfo {pages} {1} (\bibinfo {year} {2017})}\BibitemShut {NoStop}%
\bibitem [{\citenamefont {Vahlbruch}\ \emph {et~al.}(2016)\citenamefont
  {Vahlbruch}, \citenamefont {Mehmet}, \citenamefont {Danzmann},\ and\
  \citenamefont {Schnabel}}]{Vahlbruch2016}%
  \BibitemOpen
  \bibfield  {author} {\bibinfo {author} {\bibfnamefont {H.}~\bibnamefont
  {Vahlbruch}}, \bibinfo {author} {\bibfnamefont {M.}~\bibnamefont {Mehmet}},
  \bibinfo {author} {\bibfnamefont {K.}~\bibnamefont {Danzmann}},\ and\
  \bibinfo {author} {\bibfnamefont {R.}~\bibnamefont {Schnabel}},\ }\bibfield
  {title} {\bibinfo {title} {Detection of 15 {{dB}} squeezed states of light
  and their application for the absolute calibration of photoelectric quantum
  efficiency},\ }\href {https://doi.org/10.1103/PhysRevLett.117.110801}
  {\bibfield  {journal} {\bibinfo  {journal} {Physical Review Letters}\
  }\textbf {\bibinfo {volume} {117}},\ \bibinfo {pages} {110801} (\bibinfo
  {year} {2016})}\BibitemShut {NoStop}%
\bibitem [{\citenamefont {Mehmet}\ \emph {et~al.}(2011)\citenamefont {Mehmet},
  \citenamefont {Ast}, \citenamefont {Eberle}, \citenamefont {Steinlechner},
  \citenamefont {Vahlbruch},\ and\ \citenamefont {Schnabel}}]{Mehmet2011}%
  \BibitemOpen
  \bibfield  {author} {\bibinfo {author} {\bibfnamefont {M.}~\bibnamefont
  {Mehmet}}, \bibinfo {author} {\bibfnamefont {S.}~\bibnamefont {Ast}},
  \bibinfo {author} {\bibfnamefont {T.}~\bibnamefont {Eberle}}, \bibinfo
  {author} {\bibfnamefont {S.}~\bibnamefont {Steinlechner}}, \bibinfo {author}
  {\bibfnamefont {H.}~\bibnamefont {Vahlbruch}},\ and\ \bibinfo {author}
  {\bibfnamefont {R.}~\bibnamefont {Schnabel}},\ }\bibfield  {title} {\bibinfo
  {title} {Squeezed light at 1550 nm with a quantum noise reduction of 12.3
  {{dB}}.},\ }\href@noop {} {\bibfield  {journal} {\bibinfo  {journal} {Optics
  Express}\ }\textbf {\bibinfo {volume} {19}},\ \bibinfo {pages} {25763}
  (\bibinfo {year} {2011})},\ \Eprint {https://arxiv.org/abs/22273968}
  {22273968} \BibitemShut {NoStop}%
\bibitem [{\citenamefont {Tsang}(2013)}]{tsang2013quantum}%
  \BibitemOpen
  \bibfield  {author} {\bibinfo {author} {\bibfnamefont {M.}~\bibnamefont
  {Tsang}},\ }\bibfield  {title} {\bibinfo {title} {Quantum metrology with open
  dynamical systems},\ }\href@noop {} {\bibfield  {journal} {\bibinfo
  {journal} {New Journal of Physics}\ }\textbf {\bibinfo {volume} {15}},\
  \bibinfo {pages} {073005} (\bibinfo {year} {2013})}\BibitemShut {NoStop}%
\bibitem [{\citenamefont {Alipour}\ \emph {et~al.}(2014)\citenamefont
  {Alipour}, \citenamefont {Mehboudi},\ and\ \citenamefont
  {Rezakhani}}]{alipour2014quantum}%
  \BibitemOpen
  \bibfield  {author} {\bibinfo {author} {\bibfnamefont {S.}~\bibnamefont
  {Alipour}}, \bibinfo {author} {\bibfnamefont {M.}~\bibnamefont {Mehboudi}},\
  and\ \bibinfo {author} {\bibfnamefont {A.~T.}\ \bibnamefont {Rezakhani}},\
  }\bibfield  {title} {\bibinfo {title} {Quantum metrology in open systems:
  dissipative cram{\'e}r-rao bound},\ }\href@noop {} {\bibfield  {journal}
  {\bibinfo  {journal} {Physical review letters}\ }\textbf {\bibinfo {volume}
  {112}},\ \bibinfo {pages} {120405} (\bibinfo {year} {2014})}\BibitemShut
  {NoStop}%
\bibitem [{\citenamefont {Fiderer}\ \emph {et~al.}(2019)\citenamefont
  {Fiderer}, \citenamefont {Fra{\"\i}sse},\ and\ \citenamefont
  {Braun}}]{fiderer2019maximal}%
  \BibitemOpen
  \bibfield  {author} {\bibinfo {author} {\bibfnamefont {L.~J.}\ \bibnamefont
  {Fiderer}}, \bibinfo {author} {\bibfnamefont {J.~M.~E.}\ \bibnamefont
  {Fra{\"\i}sse}},\ and\ \bibinfo {author} {\bibfnamefont {D.}~\bibnamefont
  {Braun}},\ }\bibfield  {title} {\bibinfo {title} {Maximal quantum fisher
  information for mixed states},\ }\href@noop {} {\bibfield  {journal}
  {\bibinfo  {journal} {Physical review letters}\ }\textbf {\bibinfo {volume}
  {123}},\ \bibinfo {pages} {250502} (\bibinfo {year} {2019})}\BibitemShut
  {NoStop}%
\bibitem [{\citenamefont {Caves}(1981)}]{Caves1981}%
  \BibitemOpen
  \bibfield  {author} {\bibinfo {author} {\bibfnamefont {C.~M.}\ \bibnamefont
  {Caves}},\ }\bibfield  {title} {\bibinfo {title} {Quantum-mechanical noise in
  an interferometer},\ }\href
  {https://doi.org/http://dx.doi.org/10.1103/PhysRevD.23.1693} {\bibfield
  {journal} {\bibinfo  {journal} {Physical Review D}\ }\textbf {\bibinfo
  {volume} {23}},\ \bibinfo {pages} {1693} (\bibinfo {year}
  {1981})}\BibitemShut {NoStop}%
\bibitem [{\citenamefont {Colombo}\ \emph {et~al.}(2022)\citenamefont
  {Colombo}, \citenamefont {{Pedrozo-Pe{\~n}afiel}}, \citenamefont
  {Adiyatullin}, \citenamefont {Li}, \citenamefont {Mendez}, \citenamefont
  {Shu},\ and\ \citenamefont
  {Vuleti{\'c}}}]{colomboTimereversalbasedQuantumMetrology2022}%
  \BibitemOpen
  \bibfield  {author} {\bibinfo {author} {\bibfnamefont {S.}~\bibnamefont
  {Colombo}}, \bibinfo {author} {\bibfnamefont {E.}~\bibnamefont
  {{Pedrozo-Pe{\~n}afiel}}}, \bibinfo {author} {\bibfnamefont {A.~F.}\
  \bibnamefont {Adiyatullin}}, \bibinfo {author} {\bibfnamefont
  {Z.}~\bibnamefont {Li}}, \bibinfo {author} {\bibfnamefont {E.}~\bibnamefont
  {Mendez}}, \bibinfo {author} {\bibfnamefont {C.}~\bibnamefont {Shu}},\ and\
  \bibinfo {author} {\bibfnamefont {V.}~\bibnamefont {Vuleti{\'c}}},\
  }\bibfield  {title} {\bibinfo {title} {Time-reversal-based quantum metrology
  with many-body entangled states},\ }\href
  {https://doi.org/10.1038/s41567-022-01653-5} {\bibfield  {journal} {\bibinfo
  {journal} {Nature Physics}\ ,\ \bibinfo {pages} {1}} (\bibinfo {year}
  {2022})}\BibitemShut {NoStop}%
\bibitem [{\citenamefont {Manceau}\ \emph {et~al.}(2017)\citenamefont
  {Manceau}, \citenamefont {Leuchs}, \citenamefont {Khalili},\ and\
  \citenamefont {Chekhova}}]{manceauDetectionLossTolerant2017}%
  \BibitemOpen
  \bibfield  {author} {\bibinfo {author} {\bibfnamefont {M.}~\bibnamefont
  {Manceau}}, \bibinfo {author} {\bibfnamefont {G.}~\bibnamefont {Leuchs}},
  \bibinfo {author} {\bibfnamefont {F.}~\bibnamefont {Khalili}},\ and\ \bibinfo
  {author} {\bibfnamefont {M.}~\bibnamefont {Chekhova}},\ }\bibfield  {title}
  {\bibinfo {title} {Detection {{Loss Tolerant Supersensitive Phase
  Measurement}} with an {{SU}}(1,1) {{Interferometer}}},\ }\href
  {https://doi.org/10.1103/PhysRevLett.119.223604} {\bibfield  {journal}
  {\bibinfo  {journal} {Physical Review Letters}\ }\textbf {\bibinfo {volume}
  {119}},\ \bibinfo {pages} {223604} (\bibinfo {year} {2017})}\BibitemShut
  {NoStop}%
\bibitem [{\citenamefont
  {Ou}(2012)}]{ouEnhancementPhasemeasurementSensitivity2012}%
  \BibitemOpen
  \bibfield  {author} {\bibinfo {author} {\bibfnamefont {Z.~Y.}\ \bibnamefont
  {Ou}},\ }\bibfield  {title} {\bibinfo {title} {Enhancement of the
  phase-measurement sensitivity beyond the standard quantum limit by a
  nonlinear interferometer},\ }\href
  {https://doi.org/10.1103/PhysRevA.85.023815} {\bibfield  {journal} {\bibinfo
  {journal} {Physical Review A}\ }\textbf {\bibinfo {volume} {85}},\ \bibinfo
  {pages} {023815} (\bibinfo {year} {2012})}\BibitemShut {NoStop}%
\bibitem [{\citenamefont {Knyazev}\ \emph {et~al.}(2019)\citenamefont
  {Knyazev}, \citenamefont {Khalili},\ and\ \citenamefont
  {Chekhova}}]{knyazevOvercomingInefficientDetection2019}%
  \BibitemOpen
  \bibfield  {author} {\bibinfo {author} {\bibfnamefont {E.}~\bibnamefont
  {Knyazev}}, \bibinfo {author} {\bibfnamefont {F.~Y.}\ \bibnamefont
  {Khalili}},\ and\ \bibinfo {author} {\bibfnamefont {M.~V.}\ \bibnamefont
  {Chekhova}},\ }\bibfield  {title} {\bibinfo {title} {Overcoming inefficient
  detection in sub-shot-noise absorption measurement and imaging},\ }\href
  {https://doi.org/10.1364/OE.27.007868} {\bibfield  {journal} {\bibinfo
  {journal} {Optics Express}\ }\textbf {\bibinfo {volume} {27}},\ \bibinfo
  {pages} {7868} (\bibinfo {year} {2019})}\BibitemShut {NoStop}%
\bibitem [{\citenamefont {Frascella}\ \emph {et~al.}(2021)\citenamefont
  {Frascella}, \citenamefont {Agne}, \citenamefont {Khalili},\ and\
  \citenamefont {Chekhova}}]{frascellaOvercomingDetectionLoss2021}%
  \BibitemOpen
  \bibfield  {author} {\bibinfo {author} {\bibfnamefont {G.}~\bibnamefont
  {Frascella}}, \bibinfo {author} {\bibfnamefont {S.}~\bibnamefont {Agne}},
  \bibinfo {author} {\bibfnamefont {F.~Y.}\ \bibnamefont {Khalili}},\ and\
  \bibinfo {author} {\bibfnamefont {M.~V.}\ \bibnamefont {Chekhova}},\
  }\bibfield  {title} {\bibinfo {title} {Overcoming detection loss and noise in
  squeezing-based optical sensing},\ }\href
  {https://doi.org/10.1038/s41534-021-00407-0} {\bibfield  {journal} {\bibinfo
  {journal} {npj Quantum Information}\ }\textbf {\bibinfo {volume} {7}},\
  \bibinfo {pages} {1} (\bibinfo {year} {2021})}\BibitemShut {NoStop}%
\bibitem [{\citenamefont {Ramelow}\ \emph {et~al.}(2019)\citenamefont
  {Ramelow}, \citenamefont {Farsi}, \citenamefont {Vernon}, \citenamefont
  {Clemmen}, \citenamefont {Ji}, \citenamefont {Sipe}, \citenamefont
  {Liscidini}, \citenamefont {Lipson},\ and\ \citenamefont
  {Gaeta}}]{Ramelow2019}%
  \BibitemOpen
  \bibfield  {author} {\bibinfo {author} {\bibfnamefont {S.}~\bibnamefont
  {Ramelow}}, \bibinfo {author} {\bibfnamefont {A.}~\bibnamefont {Farsi}},
  \bibinfo {author} {\bibfnamefont {Z.}~\bibnamefont {Vernon}}, \bibinfo
  {author} {\bibfnamefont {S.}~\bibnamefont {Clemmen}}, \bibinfo {author}
  {\bibfnamefont {X.}~\bibnamefont {Ji}}, \bibinfo {author} {\bibfnamefont
  {J.~E.}\ \bibnamefont {Sipe}}, \bibinfo {author} {\bibfnamefont
  {M.}~\bibnamefont {Liscidini}}, \bibinfo {author} {\bibfnamefont
  {M.}~\bibnamefont {Lipson}},\ and\ \bibinfo {author} {\bibfnamefont {A.~L.}\
  \bibnamefont {Gaeta}},\ }\bibfield  {title} {\bibinfo {title} {Strong
  {{Nonlinear Coupling}} in a {{Si3N4 Ring Resonator}}},\ }\href
  {https://doi.org/10.1103/PhysRevLett.122.153906} {\bibfield  {journal}
  {\bibinfo  {journal} {Physical Review Letters}\ }\textbf {\bibinfo {volume}
  {122}},\ \bibinfo {pages} {153906} (\bibinfo {year} {2019})}\BibitemShut
  {NoStop}%
\bibitem [{\citenamefont {Strekalov}\ \emph {et~al.}(2016)\citenamefont
  {Strekalov}, \citenamefont {Marquardt}, \citenamefont {Matsko}, \citenamefont
  {Schwefel},\ and\ \citenamefont {Leuchs}}]{Strekalov2016}%
  \BibitemOpen
  \bibfield  {author} {\bibinfo {author} {\bibfnamefont {D.~V.}\ \bibnamefont
  {Strekalov}}, \bibinfo {author} {\bibfnamefont {C.}~\bibnamefont
  {Marquardt}}, \bibinfo {author} {\bibfnamefont {A.~B.}\ \bibnamefont
  {Matsko}}, \bibinfo {author} {\bibfnamefont {H.~G.~L.}\ \bibnamefont
  {Schwefel}},\ and\ \bibinfo {author} {\bibfnamefont {G.}~\bibnamefont
  {Leuchs}},\ }\bibfield  {title} {\bibinfo {title} {Nonlinear and quantum
  optics with whispering gallery resonators},\ }\href
  {https://doi.org/10.1088/2040-8978/18/12/123002} {\bibfield  {journal}
  {\bibinfo  {journal} {Journal of Optics}\ }\textbf {\bibinfo {volume} {18}},\
  \bibinfo {pages} {123002} (\bibinfo {year} {2016})}\BibitemShut {NoStop}%
\bibitem [{\citenamefont {{V.B.Braginsky, M.L.Gorodetsky,
  V.S.Ilchenko}}(1993)}]{93a1BrGoIl}%
  \BibitemOpen
  \bibfield  {author} {\bibinfo {author} {\bibnamefont {{V.B.Braginsky,
  M.L.Gorodetsky, V.S.Ilchenko}}},\ }\bibfield  {title} {\bibinfo {title}
  {Optical whispering-gallery microresonators},\ }\href@noop {} {\bibfield
  {journal} {\bibinfo  {journal} {Proceedings of SPIE}\ }\textbf {\bibinfo
  {volume} {2097}},\ \bibinfo {pages} {283} (\bibinfo {year}
  {1993})}\BibitemShut {NoStop}%
\bibitem [{\citenamefont {{M.L.Gorodetsky, V.S.Ilchenko}}(1994)}]{94a1GoIl}%
  \BibitemOpen
  \bibfield  {author} {\bibinfo {author} {\bibnamefont {{M.L.Gorodetsky,
  V.S.Ilchenko}}},\ }\bibfield  {title} {\bibinfo {title} {High-{{Q}} optical
  whispering-gallery microresonators: Precession approach for spherical mode
  analysis and emission patterns with prism couplers},\ }\href@noop {}
  {\bibfield  {journal} {\bibinfo  {journal} {Optics Communications}\ }\textbf
  {\bibinfo {volume} {113}},\ \bibinfo {pages} {133} (\bibinfo {year}
  {1994})}\BibitemShut {NoStop}%
\bibitem [{\citenamefont {Foreman}\ \emph {et~al.}(2015)\citenamefont
  {Foreman}, \citenamefont {Swaim},\ and\ \citenamefont
  {Vollmer}}]{Foreman2015}%
  \BibitemOpen
  \bibfield  {author} {\bibinfo {author} {\bibfnamefont {M.~R.}\ \bibnamefont
  {Foreman}}, \bibinfo {author} {\bibfnamefont {J.~D.}\ \bibnamefont {Swaim}},\
  and\ \bibinfo {author} {\bibfnamefont {F.}~\bibnamefont {Vollmer}},\
  }\bibfield  {title} {\bibinfo {title} {Whispering gallery mode sensors},\
  }\href {https://doi.org/10.1364/aop.7.000168} {\bibfield  {journal} {\bibinfo
   {journal} {Advances in Optics and Photonics}\ }\textbf {\bibinfo {volume}
  {7}},\ \bibinfo {pages} {168} (\bibinfo {year} {2015})}\BibitemShut {NoStop}%
\bibitem [{\citenamefont {Schliesser}\ and\ \citenamefont
  {Kippenberg}(2010)}]{Schliesser2010}%
  \BibitemOpen
  \bibfield  {author} {\bibinfo {author} {\bibfnamefont {A.}~\bibnamefont
  {Schliesser}}\ and\ \bibinfo {author} {\bibfnamefont {T.~J.}\ \bibnamefont
  {Kippenberg}},\ }\bibfield  {title} {\bibinfo {title} {Cavity optomechanics
  with whispering-gallery-mode optical micro-resonators},\ }\href@noop {}
  {\bibfield  {journal} {\bibinfo  {journal} {arXiv}\ ,\ \bibinfo {pages}
  {121}} (\bibinfo {year} {2010})},\ \Eprint {https://arxiv.org/abs/1003.5922}
  {arXiv:1003.5922} \BibitemShut {NoStop}%
\bibitem [{\citenamefont {Dmitriev}\ \emph {et~al.}(2022)\citenamefont
  {Dmitriev}, \citenamefont {Miao},\ and\ \citenamefont
  {Martynov}}]{dmitrievEnhancingSensitivityInterferometers2022}%
  \BibitemOpen
  \bibfield  {author} {\bibinfo {author} {\bibfnamefont {A.}~\bibnamefont
  {Dmitriev}}, \bibinfo {author} {\bibfnamefont {H.}~\bibnamefont {Miao}},\
  and\ \bibinfo {author} {\bibfnamefont {D.}~\bibnamefont {Martynov}},\ }\href
  {https://doi.org/10.48550/arXiv.2110.15354} {\bibinfo {title} {Enhancing the
  sensitivity of interferometers with stable phase-insensitive quantum
  filters}} (\bibinfo {year} {2022}),\ \Eprint
  {https://arxiv.org/abs/2110.15354} {arXiv:2110.15354 [gr-qc,
  physics:quant-ph]} \BibitemShut {NoStop}%
\bibitem [{\citenamefont {Gardner}\ \emph {et~al.}(2022)\citenamefont
  {Gardner}, \citenamefont {Yap}, \citenamefont {Adya}, \citenamefont {Chua},
  \citenamefont {Slagmolen},\ and\ \citenamefont
  {McClelland}}]{gardner2022nondegenerate}%
  \BibitemOpen
  \bibfield  {author} {\bibinfo {author} {\bibfnamefont {J.~W.}\ \bibnamefont
  {Gardner}}, \bibinfo {author} {\bibfnamefont {M.~J.}\ \bibnamefont {Yap}},
  \bibinfo {author} {\bibfnamefont {V.}~\bibnamefont {Adya}}, \bibinfo {author}
  {\bibfnamefont {S.}~\bibnamefont {Chua}}, \bibinfo {author} {\bibfnamefont
  {B.~J.~J.}\ \bibnamefont {Slagmolen}},\ and\ \bibinfo {author} {\bibfnamefont
  {D.~E.}\ \bibnamefont {McClelland}},\ }\bibfield  {title} {\bibinfo {title}
  {Nondegenerate internal squeezing: An all-optical, loss-resistant quantum
  technique for gravitational-wave detection},\ }\href@noop {} {\bibfield
  {journal} {\bibinfo  {journal} {Physical Review D}\ }\textbf {\bibinfo
  {volume} {106}},\ \bibinfo {pages} {L041101} (\bibinfo {year}
  {2022})}\BibitemShut {NoStop}%
\end{thebibliography}%

\end{document}